\RequirePackage[2020-02-02]{latexrelease}

\documentclass[aps,pra, twocolumn, noshowpacs, floatfix]{revtex4}

\usepackage{}
\usepackage{amsfonts}
\usepackage{amssymb}
\usepackage{graphicx}
\usepackage{amsmath}
\usepackage{amsmath}
\usepackage{braket}
\usepackage[english]{babel}
\usepackage[mathscr]{euscript}
\usepackage{color}
\usepackage{epstopdf}
\usepackage{footnote}
\usepackage{isotope}
\usepackage{ulem}

\usepackage[export]{adjustbox}

\begin{document}
\title{Dynamics of dissipative Landau-Zener transitions in an anisotropic three-level system}

\author{Lixing Zhang$^{1}$, Lu Wang$^{2}$, Maxim F. Gelin$^{3}$ and Yang Zhao$^{1}$\footnote{Electronic address:~\url{YZhao@ntu.edu.sg}}}

\affiliation{$^{1}$\mbox{School of Materials Science and Engineering, Nanyang Technological University, Singapore 639798, Singapore}\\
$^{2}$\mbox{School of Science, Inner Mongolia University of Science and Technology, Inner Mongolia 014010, China} \\
$^{3}$\mbox{School of Science, Hangzhou Dianzi University, Hangzhou 310018, China}\\
}

\begin{abstract}
We investigate the  dynamics of Landau-Zener (LZ) transitions in an anisotropic, dissipative three-level LZ model (3-LZM) using the numerically accurate multiple Davydov $\mathrm{D}_{2}$ Ansatz in the framework of the time-dependent variational principle.
It is demonstrated that a non-monotonic relationship exists between the Landau-Zener transition probability and the phonon coupling strength when the 3-LZM is driven by a linear external field. Under the influence of a periodic driving field, phonon coupling may induce peaks in contour plots of the transition probability when the magnitude of the system anisotropy matches the phonon frequency. The 3-LZM coupled to a super-Ohmic phonon bath and driven by a periodic external field exhibits periodic population dynamics in which the period and amplitude of the oscillations decrease with the bath coupling strength.
\end{abstract}

\date{\today}
\maketitle

\section{introduction}

The model of a two-level quantum system possessing an avoided crossing  driven by an external field was proposed by Landau \cite{Landau} and Zener  \cite{Zener} in 1932. This extraordinary versatile model is commonly referred to as the Landau-Zener (LZ) model. It is widely used in molecular physics\cite{m_ph1,m_ph2}, quantum optics\cite{LZ-cavity1}, chemical physics\cite{chem_physics1}, and quantum information processing \cite{QIP1, QIP2} to describe such diverse systems as molecular nanomagnets \cite{nanomagnet}, Bose-Einstein condensates (BECs)\cite{BEC}, quantum annealers \cite{quenching}, nitrogen vacancy (NV) centers in diamonds \cite{NV1,NV2}, and quantum dots in semiconductors \cite{QDot1,QDot2} (see Ref.~\cite{LZ_review} for a recent review).
The multilevel variant of the LZ problem has also gained considerable attention \cite{LZ_review}. In multilevel LZ systems,  transitions between several energy levels can occur simultaneously, adding greatly to the complexity of the system dynamics. Nevertheless, Chernyak, Sinitsin and their coworkers developed classification of the exactly solvable multilevel LZ Hamiltonians \cite{LZ-multi1,Volodya21}.  Kolovsky and Maksimov studied diabatic and adiabatic regimes of multilevel LZ transitions in the optically tilted Bose-Hubbard model \cite{Andrey}. Band and Avishai developed an analytical solution to a three-level LZ model with asymmetric tunneling between the states \cite{band}. Multilevel LZ models found applications in a variety of problems in physics and chemistry, such as BECs in multi-well traps \cite{LZ-BEC-well_trap}, trapped atomic gases \cite{at_gas}, triple quantum dots \cite{TQD1, TQD2},  large spin systems (e.g. NV centers  in diamonds) \cite{NV1,NV2} and Fe$_8$ molecular nanomagnets \cite{nanomagnet}.

A notable extension of the LZ model is achieved through the coupling of the bare LZ system to a dissipative environment.
Two-level LZ systems diagonally and off-diagonally coupled to phonon baths have been extensively studied  \cite{LZ_review}. It was shown, for example,  that  phonon coupling may  assist  LZ transitions and create entanglement between qubits \cite{D2-dw3,LZ-periodic2, LZ-entanglement1,LZ-entanglement2}.
Saito, Wurb and their coworkers derived exact analytical formula for the transition probability in the dissipative LZ model \cite{bath1,bath2} which was  confirmed by accurate numerical calculations \cite{YZ19}.
Nalbach and Thorwart gave numerically-exact path-integral simulations of the dissipative LZ dynamics \cite{Thorwart09,Thorwart15}.
As for multilevel LZ models, the studies of  dissipative effects are scarce. Mention the work of
Saito {\it et al.} who investigated dissipative LZ transitions by employing diabatic representation of the LZ Hamiltonian \cite{dibatic_basis}.

In the present work, we study dynamics of the dissipative  three-level LZ model (hereafter, 3-LZM). We
 consider an anisotropic 3-LZM, similar to that of Ref.~\cite{Core-kiselev}, in which degeneracy between the three energy levels is lifted (without the  anisotropy term, the 3-LZM is a direct extension of the original two-level LZ model \cite{extention1, extention2}. Anisotropic 3-LZMs are commonly used to simulate NV centers  in diamonds \cite{NV_review1} and triple quantum dots \cite{TQD1, TQD2},  quantum computers  \cite{NV-computing1, NV-computing2, NV-computing3}, stress sensors \cite{NV-stress}, quantum cellular automata \cite{TQD-automata}, and charging rectifiers \cite{TQD-rectifier}.

Evaluation of the  dissipative driven 3-LZM dynamics is computationally challenging. 
It is not surprising therefore that the influence of phonon modes  on the 3-LZM dynamics was studied so far without external driving and  either for a single harmonic mode  \cite{Ashhab16} or in terms of phenomenological Markovian master equations \cite{Militello19a,Militello19b,Militello19c}. Our goal is much more ambitious. We aim to develop a methodology for the numerically accurate simulation of 3-LZMs  driven by external fields of any shape and coupled to bosonic baths with arbitrary spectral densities. In this work, we tackle this problem by invoking the numerically accurate method of the multiple Davydov D$_2$ (multi-D$_2$) Ansatz, which is a member of the family of variational Gaussian Davydov Ans\"atze \cite{D2-review}. The multi-D$_2$ Ansatz accounts for bosonic environment through an expansion of the total (system + bath) many-body wave function with multiple coherent states per bath mode. The so-constructed wave function converges to a numerically ``exact
" solution to the multidimensional, many-body Schr\"odinger equation if multiplicity of the Ansatz is sufficiently high \cite{D2-review}.
The multi-D$_2$ Ansatz was demonstrated to be an efficient and reliable tool for obtaining time evolutions of photon-assisted LZ models \cite{LZ-periodic2}, the Holstein-Tavis-Cummings (HTC) model \cite{D2-HTC,D2-TC}, the spin-boson model (SBM) and its variants\cite{D2-SBM1,D2-SBM2}, and the cavity quantum electrodynamics (QED)  models \cite{D2-SF}.

The remainder of the paper is arranged as follows: In Sec.~II, we introduce the theoretical framework of the multi-D$_2$ Ansatz, as well as define the observables used in this work. In Sec.~III, we present and discuss dynamic simulations of the dissipative 3-LZM in different regimes. Sec.~IV is the Conclusion. Technical derivations can be found in Appendix A.

\section{METHODOLOGY}

\subsection{The model}
The bare 3-LZM Hamiltonian can be written as
\begin{eqnarray}\label{EQ1}
\hat{H}_{{\rm 3} - {\rm LZM}}& =& \mathcal{D}(S_z^2-\frac{2}{3}I)+ \Omega_z(t) S_z+\sqrt{2}~\Omega_x(t) S_x
\end{eqnarray}
($\hbar$=$1$) where  $S_{x}$ and $S_z$ are the spin 1 operators. The zero-field splitting (ZFS) tensor $\mathcal{D}(S_z^2-\frac{2}{3}I)$ is responsible for the system anisotropy and describes the level splitting without external fields. The  $-2I/3$ term makes the ZFS tensor traceless and restores the SU(3) symmetry of the Hamiltonian. $\Omega_z(t)$ and $\Omega_x(t)$ are time-dependent  driving fields in the $z$ and $x$ directions, which can change either linearly or periodically with time.

\begin{figure}[t]
\centerline{\includegraphics[width=80mm]{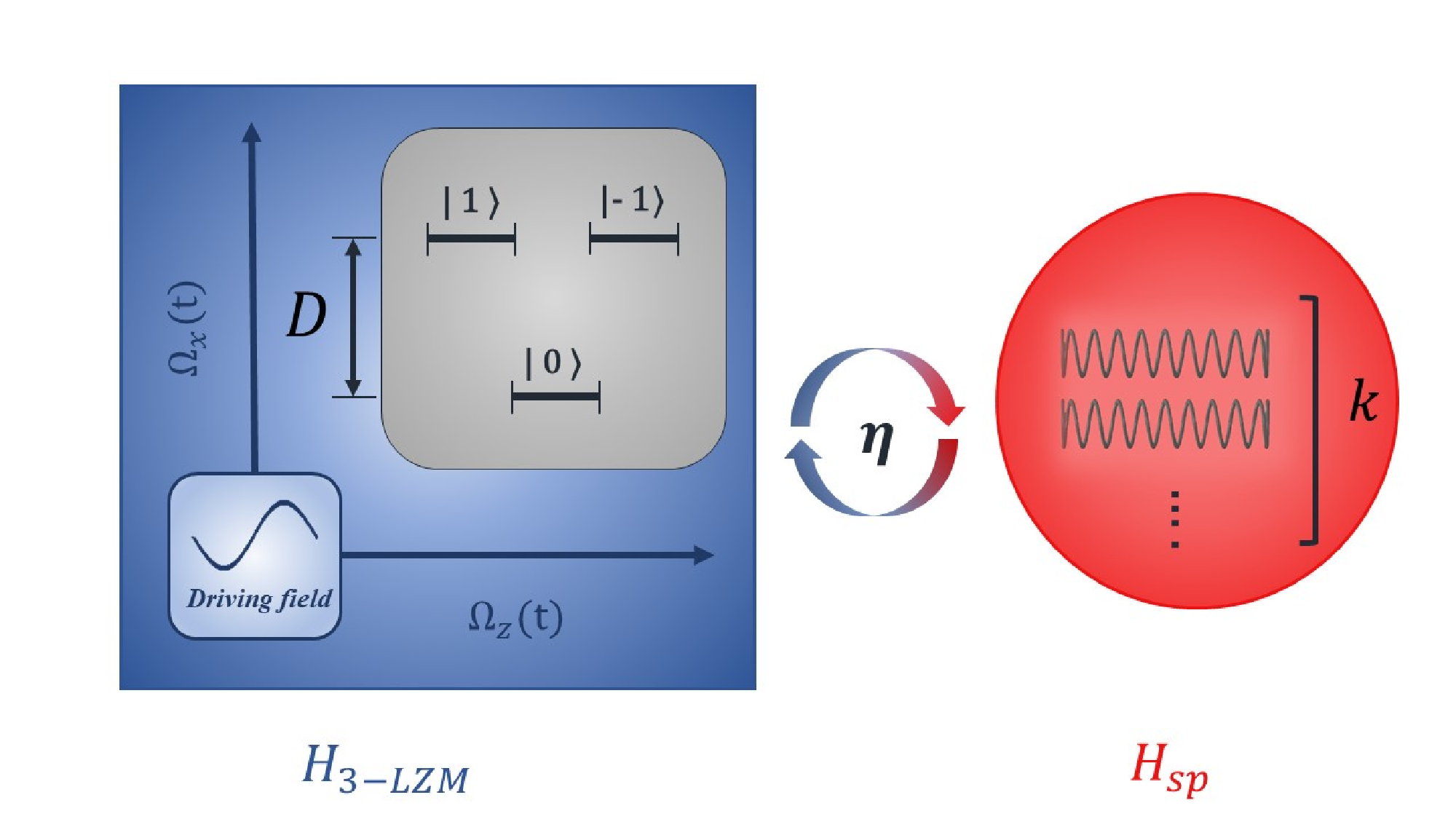}}
\caption{Schematic of the Hamiltonians $\hat{H}_{\rm 3-LZM}$ and $\hat{H}_{\rm sp}$ defined by Eqs.~(\ref{EQ1}) and (\ref{EQ2}), respectively.}
\label{Schematic}
\end{figure}

The $\hat{H}_{\rm 3-LZM}$ Hamiltonian is linearly coupled  to a phonon bath which assists transitions between the spin states, create complex interference patterns and drives the system towards certain steady states.  The spin-phonon coupling, together with the phonon bath, are described by the Hamiltonian $\hat{H}_{\rm sp}$:
\begin{eqnarray}\label{EQ2}
\hat{H}_{\rm sp}& =& \sum_{k}\omega_{ k} \hat{b}_{ k}^{\dagger}\hat{b}_{k}+\sum_{k}(\eta_z^k S_z+\eta_x^k S_x)(\hat{b}_{ k}^{\dagger}+\hat{b}_{k})
\end{eqnarray}
Here $\hat{b}_{ k}^{\dagger}$($\hat{b}_{ k}$) is the annihilation (creation) operator of the $k$th phonon mode, while  $\omega_k$, $\eta_z^k$ and $\eta_x^k$ are the frequency, diagonal coupling, and off-diagonal coupling strengths of the $k$th mode, respectively. {The cartoon view of the total system Hamiltonian can be found in Fig.~\ref{Schematic}}.

If $\hat{H}_{\rm sp}$ contains only one phonon mode, we drop the subscript $k$ and denote $\eta_z^k = \eta_z$, $\eta_x^k= \eta_x$, and $\omega_k = \omega_p$. If multiple modes are present, the  phonon bath is specified by the spectral density
\begin{equation}\label{EQ3}
J(\omega )=\sum_{k}(\eta_z^{k})^{2}\delta(\omega -\omega _{k})=2\alpha \omega_{c}^{1-s}\omega^s e^{-\omega/\omega_c}
\end{equation}
where $\alpha$ is the dimensionless coupling strength and $s$ is the exponent. When $s = 1$, the bath is Ohmic, while $s > 1$ ($s < 1$) correspond to  super-Ohmic (sub-Ohmic) bath.  Following Ref.~\cite{Core-spec_den}, we choose a super-Ohmic bath with $s = 3$ in our simulations. $\omega_{c}$ is the cut-off frequency, the coupling strength of the phonon modes with $\omega > \omega_{c}$ will decrease drastically.
In order to employ the variational Davydov Ansatz method, the continuum spectral density has to be discretized. Following  Ref.~{\cite{WFCZ}}, we employ the density function of the phonon modes ${\mathcal{P}}(\omega)$, which is defined on the interval $\omega \in [0, \omega_m]$, where $\omega_m$ is the upper bound of the phonon frequencies and
\begin{equation}\label{EQ4}
\int_{0}^{\omega_k}{\mathcal{P}}(\omega) d\omega = k, k = 1, 2, ..., N_b,
\end{equation}
where $N_b$ is the total number of discrete modes in the Hamiltonian and  $\omega_{N_b} = \omega_m$. In this case, the coupling strength of the individual phonon mode is given by the expression  $\eta_z^k= \sqrt{J(\omega_k)/\mathcal{P}(\omega_k))}$. In this work, we adopt the efficient
\begin{equation}\label{EQ5}
{\mathcal{P}}(\omega) = \frac{1}{\mathcal{N}}\frac{J(\omega)}{\omega}.
\end{equation}
Here
\begin{equation}\label{EQ6}
	{\mathcal{N}} = \frac{1}{N_b}\int_0^{\omega_m}\frac{J(\omega)}{\omega}d\omega,
\end{equation}
is the normalization factor that ensures $\int_{0}^{\omega_m}{\mathcal{P}}(\omega) d\omega = N_b$.
In the case of super-Ohmic bath with $s=3$, $\omega_k$ can only be obtained implicitly through Eq.~(\ref{EQ4}).

\subsection{The multi-D$_2$ Ansatz}
First proposed in the 1970s in the study of energy transportation in protein molecules as an approximation of the solution to the Schr\"{o}dinger equation~\cite{Davydov1}, the Davydov Ansatz has two variants, the D$_1$  and D$_2$ Ans\"atze, with the latter being a simplified variant of the former~\cite{Davydov2, Davydov4}. Inspired by numerically ``exact" solutions to a two-site problem with short-range coupling to Einstein phonons \cite{shor73}, Shore and Sander pioneered the multiple Davydov Ans\"{a}tze when they
experimented with a trial wave function with two Gaussians for the phonon component, an early forerunner of the multiple Davydov Ans\"{a}tze.
In a further development of this method, the multiple Davydov Ans\"{a}tze have been applied extensively to a variety of many-body problems in physics and chemistry by Zhao and co-workers \cite{jcppersp,Davydov5,Davydov6,Davydov7,Davydov8,QED_Huang,QED_Zheng,TMD}. The multiple Davydov Ans\"{a}tze also have two variants, multi-D$_1$ Ansatz and multi-D$_2$ Ansatz (see \cite{Perspetive} for  a recent review). In this work, the multi-D$_2$ Ansatz is used:
\begin{align}\label{EQ7}
\ket{{\rm D}_{2}^{M}(t)}=\sum_{n=1}^{M}{(A_n(t){\ket{-1}}+B_n(t){\ket{0}}+C_n(t){|1\rangle})\otimes}|f\rangle_{\rm ph}^n
\end{align}
where $M$ is the multiplicity of the Ansatz, and ${\ket{-1}}$, ${|0\rangle}$ and ${|1\rangle}$ are the three spin states in the 3-LZM, for which we assign amplitudes $A_n(t)$, $B_n(t)$ and $C_n(t)$, respectively. $|f\rangle_{\rm ph}^n$ is the phonon coherent state, which can be written as:
\begin{eqnarray}\label{EQ8}
|f\rangle_{\rm ph}^n= \exp \big[{\sum_{k} {f}_{nk}\left(t\right)\hat{b}^{\dagger}-\rm H.c.}\big]|0\rangle_{\rm ph}
\end{eqnarray}
Here, $|0\rangle_{\rm ph}$ is the phonon vacuum state,  $f_{nk} (t)$ is the phonon displacement for the $k$th mode of phonon, and H.c.~denotes the Hermitian conjugate. $A_n(t)$, $B_n(t)$, $C_n(t)$, and $f_{nk}(t)$ are referred to as the time-dependent variational parameters that define the multi-D$_2$ trial state.

\subsection{The time-dependent variational principle}
To arrive at the time-dependent variational parameters, the Euler-Lagrange equation is solved under the framework of the time-dependent variational principle:
\begin{align}\label{EQ9}
\frac{d}{dt}\frac{\partial L}{\partial{\dot{u}}_n^\ast}-\frac{\partial L}{\partial u_n^\ast}=0
\end{align}
where $u_n$ are the time-dependent variational parameters, $u_n^\ast$ stands for the complex conjugate of $u_n$, and the Lagrangian is given by:
\begin{eqnarray}\label{EQ10}
L&=&\frac{i}{2}\left[\langle {\rm D}_{2}^{M}(t)|\frac{\overrightarrow{\partial}}{\partial t}|{\rm D}_{2}^{M}(t)\rangle
-\langle {\rm D}_{2}^{M}(t)|\frac{\overleftarrow{\partial}}{\partial t}|{\rm D}_{2}^{M}(t)\rangle\right]\nonumber\\&&-\langle{\rm D}_{2}^{M}(t)|\hat{H}_{{\rm 3} - {\rm LZM}}+\hat{H}_{\rm sp}|{\rm D}_{2}^{M}(t)\rangle.
\end{eqnarray}
By solving the Euler-Lagrange equation, a series of differential equations that govern the time evolution of the variational parameters $u_n$ are obtained. These equations are referred to as the equations of motions (EOMs). To solve the EOMs simultaneously, the $4^{\rm th}$-order Runge-Kutta method (RK4) is adopted. The detailed derivation of the EOMs can be found in Appendix \ref{Appendix A}. To avoid singularity in RK4 iterations, a random noise is added to the initial variational parameters at $t=0$ within a range of $[{-10}^{-4}, {10}^{-4}]$ (see Ref.  \cite{D2-review} for the discussion of technical details).

\subsection{Observables}

Following Eq.~(\ref{EQ7}), the normalization factor of the Ansatz state can be calculated as follows:
\begin{eqnarray}\label{EQ11}
N(t) & = & \langle{\rm D}_{2}^{M}(t)|{\rm D}_{2}^{M}(t)\rangle\nonumber\\
& = & \sum_{m,n}^{M}{\left({A_m}^\ast{A_n}+{B_m}^\ast{B_n}+{C_m}^\ast{C_n}\right)\cdot S_{mn}}.
\end{eqnarray}
where
\begin{eqnarray}\label{EQ12}
S_{mn}=\exp \left[\sum_{k}({-\frac{1}{2}{f_{mk}^\ast f_{mk}-\frac{1}{2}f_{nk}^\ast f_{nk}+}f_{mk}^\ast f_{nk}})\right]
\end{eqnarray}
is the well-known Debye-Waller factor  \cite{D2-dw1, D2-dw2, D2-dw4, D2-dw5}. If the calculation converges, the normalization factor is $t$-independent \cite{D2-norm, D2-SBM2}.

The time-dependent probabilities to detect the spin in the states $\ket{0}$, $\ket{1}$ and $\ket{-1}$ are defined as
\begin{equation}\label{EQ13}
  P_{0}(t)=\braket{{\rm D}_{2}^{M}(t)|0} \braket{0|{\rm D}_{2}^{M}(t)}
=\sum_ {m,n}^{M}{\left({B^\ast_m} B_n\right)S_{mn}},
\end{equation}
\begin{equation}\label{EQ14}
P_{-1}(t)=\braket{{\rm D}_{2}^{M}(t)|-1} \braket{-1|{\rm D}_{2}^{M}(t)}
=\sum_ {m,n}^{M}{\left({A^\ast_m} A_n\right)S_{mn}},
\end{equation}
and
\begin{equation}\label{EQ15}
  P_{1}(t)=\braket{{\rm D}_{2}^{M}(t)|1} \braket{1|{\rm D}_{2}^{M}(t)}
=\sum_ {m,n}^{M}{\left({C^\ast_m} C_n\right)S_{mn}}.
\end{equation}
If  $\ket{0}$ is chosen as the initial state (this is so in the present work), then $P_{0}(t)$, $P_{1}(t)$ and $P_{-1}(t)$ express the transition probabilities to evolve from the initial state at $t=0$ to the states $\ket{0}$, $\ket{1}$ and $\ket{-1}$ at time $t$. The three probabilities satisfy the equality
\begin{equation}\label{EQ16}
P_{-1} + P_{0} + P_{1} =1,
\end{equation}
which means the possible measurement outcomes are restricted to only $\ket{-1}$, $\ket{0}$ and $\ket{1}$.


To quantify the role of the bath on the 3-LZM dynamics, we also calculate the partial populations $P_{k,n}$ of the $\ket{k}$ spin states ($k=-1$, $0$, $1$) with phonons in Fock state $\ket{n}$.
With the projection operators $\hat{P}_{k,n}=\ket{k}\bra{k}\otimes\ket{n}\bra{n}\equiv \ket{k,\ n}\bra{k,\ n}$, the partial  populations read
\begin{equation}
P_{k,n} = \mathrm{Tr} (\rho \hat{P}_{k,n}) = \left|\braket{k,\ n|\mathrm{D}_{2}^{M}} \right|^{2}
\end{equation}

\section{RESULTS AND DISCUSSION}

The 3-LZM without coupling to the phonon bath has been extensively studied \cite{Core-Statis, Core-kiselev, three1, three2,three3}, and possibility to control transitions in 3-LZM via external fields was also demonstrated theoretically \cite{Qin16}. When the driving field is linear, the  anisotropy lifts the energy difference between the $|\pm 1\rangle$ and $|0\rangle$ states. The isotropic 3-LZM ($\mathcal{D}=0$) was discussed in some detail ~\cite{three1}. When $\mathcal{D}$ is zero, transitions between the states  $|0\rangle$ and $|1\rangle$  occur simultaneously with transitions between the states $|0\rangle$ and $\ket{-1}$.  When $\mathcal{D}$ is small, the two LZ transitions are slightly separated in time, which allows for interference and yields pulse-shaped pattern in the time evolution of the transition probability. When $\mathcal{D}$ is large, the time intervals between the two LZ transitions are too long  for the interference to occur, and the transition probability dynamic resembles two sequential  LZ transitions.
When the 3-LZM is driven by the periodic external field,
$P_{-1}(t) = P_{1}(t)$ owing to the symmetry of the Hamiltonian. In this case, $P_{\pm 1}(t)$ behaves non-trivially  if the driving field  and anisotropy term fulfill   appropriate resonance conditions as discussed in Ref.~\cite{Core-Statis}. The origin of such phenomenon can be attributed to the coherent destruction of tunneling (CDT) \cite{CDT}, which is responsible for the disappearance of tunneling effect in some parameter regimes.

In this section, we study how the above results are modified if the 3-LZM coupled to  phonon modes. In sections III.A and III.B, we consider, respectively,  linearly and periodically driven LZM-3 models coupled to a single harmonic mode.  In section III.C, we explore how a multimode phonon bath affects the LZM-3 dynamics. Note that  $\omega_z$ is chosen as the unit frequency $\omega$ in this section.

\subsection{Single-mode phonon coupling with linear driving field}

In this subsection, we study a dissipative 3-LZM driven by an external field linearly, that is for  $\Omega_{z}(t)=vt$. Here $v$ is the scanning velocity which describes, e.g.,  the changing speed of the external magnetic field. The tunneling rate between the adjacent states, $\Omega_{x}(t)$, is set to $\Delta = {\rm const}$. ZFS $\mathcal{D}$ is fixed at $10$ due to the  following consideration. If a 3-LZM does not interact with a phonon bath, crossing of the energy levels of $\ket{1}$, $\ket{0}$, $\ket{-1}$ is determined by $\mathcal{D}$. Namely, intersection points between $\ket{1}$ and $\ket{0}$, $\ket{1}$ and $\ket{-1}$ as well as $\ket{-1}$ and $\ket{0}$ are separated by  $2\mathcal{D}$ \cite{Core-kiselev}. If $\mathcal{D}$ is small, $\ket{1}$ and $\ket{0}$ intersect at $vt=-\mathcal{D}$, $\ket{1}$ and $\ket{-1}$ at $t=0$, and $\ket{-1}$ and $\ket{0}$ at $vt=\mathcal{D}$. The avoided crossings at these close intersecting points will ``interfere'' with each other. In order to avoid the complex interference at small anisotropy, $\mathcal{D}=10$ is selected. In all calculations of this subsection, multiplicity  $M = 6$ of the $D_2$ Ansatz is used for obtaining converged results.

\begin{figure}[t]
\centerline{\includegraphics[width=80mm]{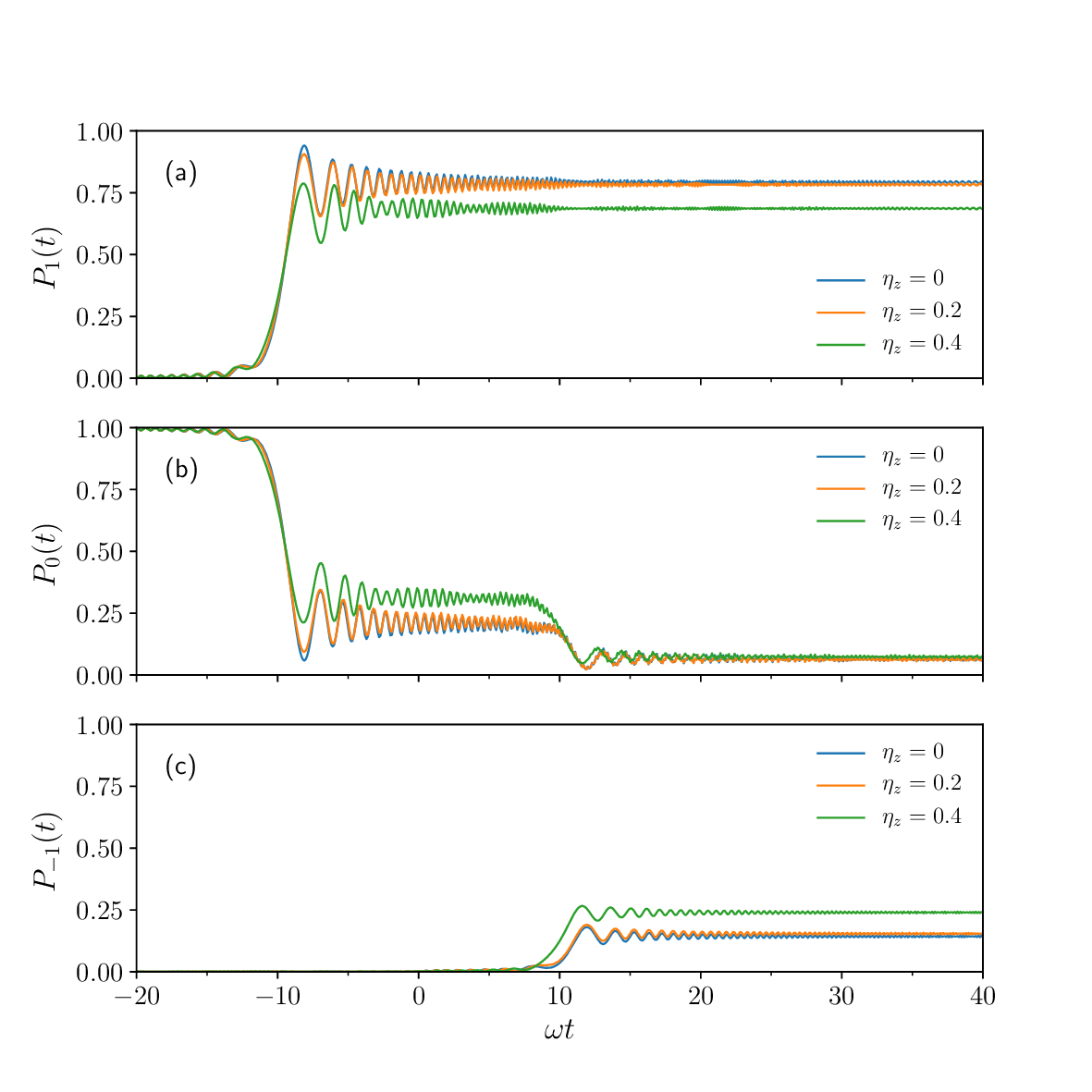}}
\centerline{\includegraphics[width=80mm]{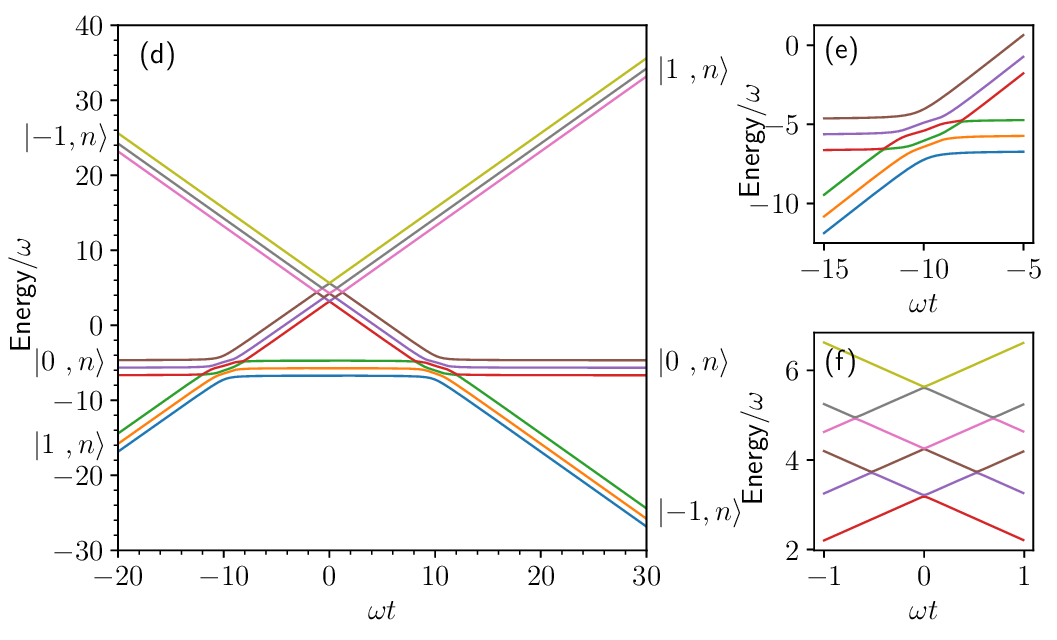}}
\caption{Panels (a),(b) and (c): Time evolution of the LZ transition probabilities $P_{1} (t)$, $P_{0} (t)$ and $P_{-1}(t)$, respectively, from $\omega t = -20$ to $\omega t = 40$ for $\eta_z=0$ (blue line), $\eta_z/\omega=0.2 $ (orange line), and $\eta_z/\omega=0.4 $ (green line). The remaining  parameters are as follows: $v/\omega^2 = 1 $, $\mathcal{D} /\omega= 10$, $\Delta /\omega= 0.5$, and $\omega_{p}/\omega = 1$. Panel (d): The corresponding energy diagram for the Hamiltonian with the same parameters, except for $\eta_{z} = 0.4\omega$. The Fock state is restricted to an upper limit of $|{2}\rangle$. Details of panel (d) at $\omega t=-10$ and $0$ are amplified in panels  (e) and (f), respectively. }
\label{Fig1}
\end{figure}

{We first investigate how the diagonal coupling strength $\eta_{z}$ affects the dynamics of the LZ transition. The time dependent probabilities $P_{1} (t)$, $P_{0} (t)$ and $P_{-1}(t)$ are plotted in Figs.~\ref{Fig1} (a), (b) and (c), respectively, for three values of $\eta_{z}$.
The energy levels
involved in the LZ dynamics are shown in Fig.~\ref{Fig1} (d).
Details of Fig.~\ref{Fig1} (d) at $\omega t=-10$ and $0$ are amplified in Figs.~\ref{Fig1} (e) and (f), respectively. As shown
in Fig.~\ref{Fig1} (a), $P_{1} (t)$ rises at $\omega t=-\mathcal{D}\omega/v=-10$ where a LZ transition occurs between $\ket{1}$ and $\ket{0}$. In Fig.~\ref{Fig1}
(d), one can see that $\ket{1}$ increases with time while $\ket{0}$ remains constant under the linearly changing field ($vt$).
It follows that $\ket{1}$ and $\ket{0}$ meet at
$\omega t=-\mathcal{D}\omega/v=-10$. At $\omega t=-\mathcal{D}\omega/v=-10$,
a corresponding decrease in $P_{0}$ can be seen in Fig.~\ref{Fig1} (b). Meanwhile, the gap is open between
$\ket{1,n}$ and $\ket{0,m}$ (see Fig.~\ref{Fig1} (e)), where $n$ and $m$ denote the photonic Fock states.
Similarly, $P_{0} (t)$ dips at $\omega t=\mathcal{D}\omega/v=10$ in Fig.~\ref{Fig1} (b), and the corresponding increase in $P_{-1} (t)$ can be seen in Fig.~\ref{Fig1} (c). Both events correspond to the LZ transition between $\ket{0}$ and $\ket{-1}$, which is shown in Fig.~\ref{Fig1} (d).

According to Fig.~\ref{Fig1} (b),  $P_{0} (t)$ decreases in the vicinity of $ \omega t=\mathcal{D}\omega/v=10 $ with a simultaneous rise of $P_{-1}(t)$ (see  Fig.~\ref{Fig1} (c)), which corresponds to the avoided crossing between $\ket{-1}$ and $\ket{0}$.
In Fig.~\ref{Fig1}, the curves for $\eta_{z}/\omega=0$ and $0.2$ nearly overlap, while the curve for $\eta_{z}/\omega=0.4$ deviates substantially from them.  This indicates that $\eta_{z}/\omega=0.4$ is a relative strong coupling strength for the present model, while $\eta_{z}/\omega=0.2$ is relatively weak. How exactly the coupling strength $\eta_{z}$ influences the dynamics is a tough question. As follows from Fig.~\ref{Fig1} (b), quenching of $P_{0} (t)$ of $\eta_{z}/\omega=0.4$ at $\omega t=10$ ($-10$) is faster (slower) than those of $\eta_{z}/\omega=0$ and 0.2. Correspondingly, $P_{1} (t)$ ($P_{-1} (t)$) of $\eta_{z}/\omega=0.4$ rises faster (slower) at $\omega t=10$ ($-10 $) in Fig.~\ref{Fig1} (a) (Fig.~\ref{Fig1} (c)). At $t=0$, there is no gap as shown in Fig.~\ref{Fig1} (f), and there is no transition between $\ket{-1}$ and $\ket{1}$. From Fig.~\ref{Fig1} (d), one can see that the energy levels at $\omega t=-10$ and $\omega t=10$ are symmetric. However, the decay rates of $P_{0} (t)$ for $\eta_{z}/\omega=0.4$ at $\omega t=-10$ and $\omega t=10$ are not the same.}

\begin{figure}[t]
\centerline{\includegraphics[width=85mm,scale=1]{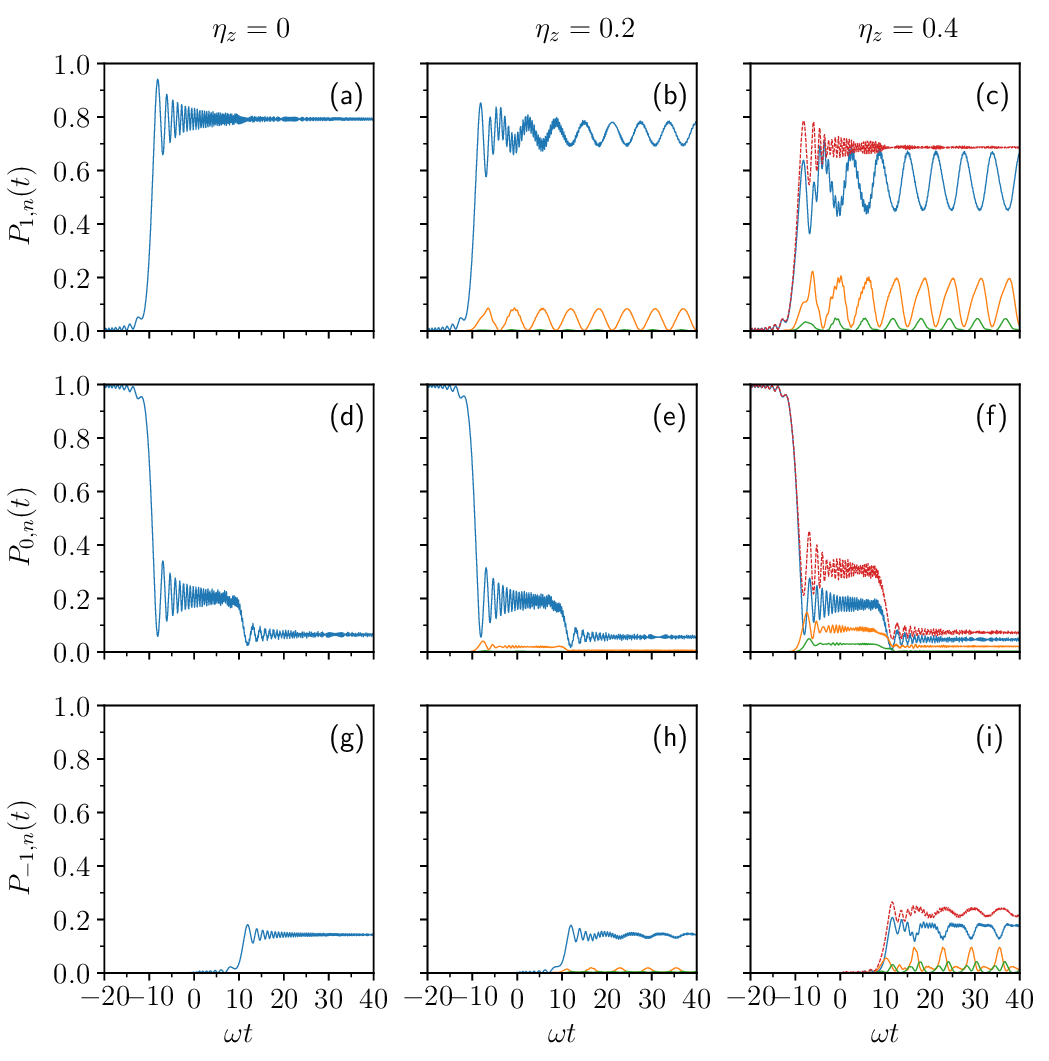}}
\caption{Time evolution of the population $P_{k,n}(t)$ (from $\omega t = -20$ to $\omega t = 40 $). $P_{1,n}$ are plotted in panels (a), (b) and (c);
$P_{0,n}$ are plotted in panels (d), (e) and (f);  $P_{-1,n}$ are plotted in panels (g), (h) and (i). $\eta_{z}/\omega=0$ corresponds to panels (a), (d) and (g); $\eta_{z}/\omega=0.2$ corresponds to panels (b), (e) and (h);  $\eta_{z}/\omega=0.4$ corresponds to panels (c), (f) and (i). Other parameters are the same as those in Fig.~\ref{Fig1}. $P_{k,0}(t)$:  blue lines; $P_{k,1}(t)$: orange lines; $P_{k,2}(t)$: green lines; Total $P_{k,n}(t)=\sum_{n=0}^{2}P_{0,n}(t)$: red dashed lines in panels (c), (f) and (i). }
\label{Fig2}
\end{figure}

To understand the asymmetry of the dynamics at $\omega t=-10$ and $10$, populations $P_{k,n}$ are calculated and plotted in Fig.~\ref{Fig2}. The figures are arranged as follows. Curves with the same spin are placed in the same row. $P_{1,n}$ are plotted in Figs.~\ref{Fig2} (a), (b) and (c), $P_{0,n}$ are plotted in Figs.~\ref{Fig2} (d), (e) and (f), and $P_{-1,n}$ are plotted in Figs.~\ref{Fig2} (g), (h) and (i). Curves with the same coupling strength are placed in the same column. $P_{k,n}(t)$ of $\eta_z/\omega=0$ are in Figs.~\ref{Fig2} (a), (d) and (g), $P_{k,n}(t)$ of $\eta_z/\omega=0.2$ are in Figs.~\ref{Fig2} (b), (e) and (h), and $P_{k,n}(t)$ of $\eta_z/\omega=0.4$ are in Figs.~\ref{Fig2} (c), (f), (i).

The complex dynamic behavior of $P_{z}(t)$ can be understood by the inspection of $P_{k,n}(t)$, which are presented in Figs.~\ref{Fig2} (a)-(i)  for several Fock states $\ket{n}$:  $P_{k,0}$ are blue, $P_{k,1}$ are orange, and $P_{k,2}$ are green. For $\eta_z/\omega=0$ ($\eta_z/\omega=0.2$), $P_{k,n}(t)$ for $n\geqslant 1$ ($n\geqslant 2$) are negligibly small and invisible. For $P_{k,n}(t)$ and $\eta_z/\omega=0$, the populations in Figs.~\ref{Fig2} (a), (d) and (g)  are similar to $P_{z}(t)$ in Figs.~\ref{Fig1} (a), (b) and (c), respectively, which lends support to our analysis. The most prominent difference of $P_{k,n}(t)$ for $\eta_z/\omega=0.4$ with those of $\eta_z/\omega=0$ and 0.2 is as follows: $P_{k,n}(t)$ with $n\geqslant 1$ are much larger. Hence the evolution of  $P_{z}(t)$ is influenced not only by the vacuum state $\ket {n = {0}}$ of the bath, but also by the Fock states with $n\geqslant 1$. The dynamics of $P_{z}(t)$ reflects the total impact of both ground and excited states of the bath. According to Fig.~\ref{Fig2} (f), the magnitude  of $P_{0, 0}(t)$ is the same as those in Fig.~\ref{Fig2} (d) and (e). However, $P_{0,n=1,2}(t)$ are much larger. We plot the total populations $\sum_{n=0}^{2}P_{0,n} (t)$ with red dashed lines in Figs.~\ref{Fig2} (c), (f) and (i). One can easily see that this sum  Fig.~\ref{Fig2} (f) is larger than $P_{0,0} (t) $ in Figs.~\ref{Fig2} (d) and (e).
The total populations  $\sum_{n=0}^{2}P_{k,n} (t)$ for $k= \pm 1$ in Figs.~\ref{Fig2} (c) and (i) also support our above analysis. Seen alone, the complex spin dynamics $P_{z}(t)$ may be difficult to understand. The detailed dynamics projected on the Fock states of the bath, as revealed by our variational method, provides great help in deciphering the spin dynamics.

Apart from the dynamic asymmetry,  Fig.~\ref{Fig2} reveals vividly how coupling to the vibrational mode modifies the 3-LZM interference patterns. Namely, on top of  standard high-frequency oscillations with decreasing period and amplitude (which originate from the wave-packet re-crossings) we observe purely vibrational oscillations with a period $\sim 2\pi/\omega = 2 \pi$. These latter  oscillations are absent in the first column of Fig.~\ref{Fig2} (an isolated 3-LZM system), but are clearly visible in panels (b), (c) and (f) (3-LZM coupled to the vibrational mode). Vibrational features are especially pronounced in partial populations, and the difference in shape of vibrational beatings in  $P_{1,n}(t)$ and $P_{-1,n}(t)$ can be attributed to the dynamic asymmetry discussed above.
For specific $k=1$ or $-1$, vibrational oscillations  in $P_{k,n}(t)$ are shifted with respect to each other for different $n$.
As the result of this shift, the influence of the vibrational modes on the total populations $P_{1}(t)$ and $P_{-1}(t)$ is milder. Yet, pronounced $2 \pi$-periodic vibrational modulations of the high-frequency low-amplitude oscillations are clearly seen in panels (c) and (i). Similar vibrational modulations were also revealed in the 2-LZM coupled to a harmonic oscillator  \cite{Malla18}.

\subsection{Single-mode phonon coupling with periodic driving field}

We now switch to the study of the 3-LZM in the presence of the periodic driving field. We set $\Omega_z(t) = A_{\rm z}\cos(\omega_{z} t)$ and $\Omega_{x}(t) = A_{x}\cos(\omega_{x} t)$, where $A_{z}$($A_{x}$) and $\omega_{z}$($\omega_{x}$) are the amplitude and frequency of the driving field in $z$ ($x$) direction, respectively. As has already been mentioned, $P_{- 1}(t) = P_{1}(t)$ for periodic driving field. If one of the three populations is known, one can immediately obtain the other two from Eq.~(\ref{EQ16}). We choose to study the effects of periodic driving on $P_{-1}(t)$.  In the Hamiltonian of Eq.~(\ref{EQ2}), the phonon bath  can couple the 3-LZM in both $x$ and $z$ directions. Rather trivial changes in the dynamics are found in the case of $z$-coupling. Hence we will focus on  the $x$-coupling. In all calculations of this subsection, multiplicity  $M = 6$ of the $D_2$ Ansatz is used for obtaining converged results.

\begin{figure}[h]
\adjustimage{width=.35\textwidth,center}{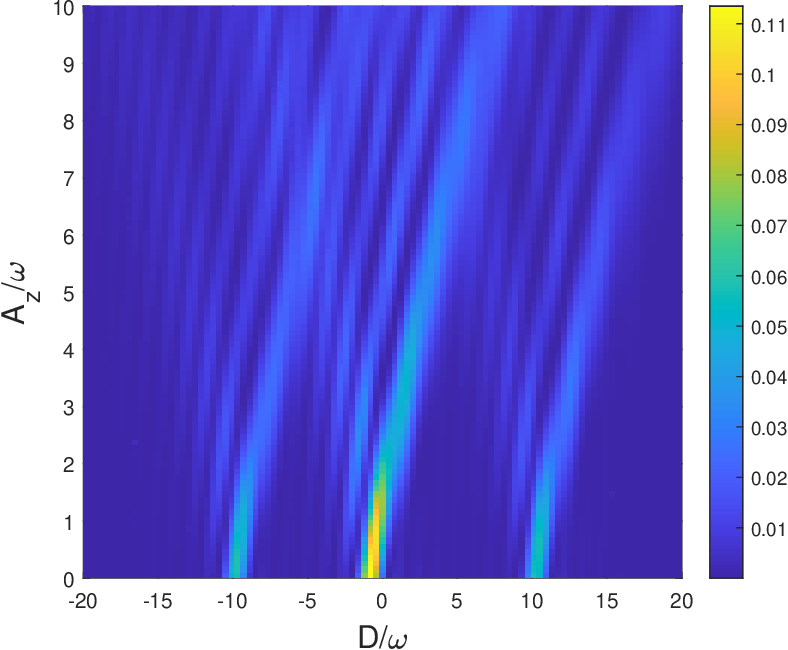}
\caption{Contour plot of $P_{-1}( t = 5/\omega)$ as a function of $A_z/\omega$ and $\mathcal{D}/\omega$. The $x$-direction driving field amplitude $A_x /\omega = 0.1$, and $\omega_{x} /\omega= 10$. The phonon frequency $\omega_{p} /\omega= 1$, the $x$-direction phonon coupling strength $\eta_{x} /\omega= 0.1$, and the $z$-direction phonon coupling strength $\eta_{z} = 0$. }
\label{Fig3}
\end{figure}

Following  \cite{Core-contour_plot}, we use contour plots for visualizing the periodic driving effects. The contour plot
 of $P_{-1}( t = 5 /\omega)$ is displayed in Fig.~\ref{Fig3} as a function of the anisotropy $\mathcal{D}$ and  the  amplitude of the driving field $A_z$. The remaining parameters are as follows: $\omega_x /\omega= 10$, $A_x/\omega = 0.1$ and the phonon frequency $\omega_{p}/\omega = 1$. The phonon mode is coupled to the 3-LZM in the $x$-direction with a coupling strength $\eta_{x}/\omega = 0.1$. The  driving field amplitude in the $x$-direction is chosen to be $A_x /\omega = 0.1$.

Three main peaks appear in Fig.~\ref{Fig3} for weak driving amplitudes. Based on their origins, the peaks can be grouped as follows: (i) a pair of peaks that are centered at $\mathcal{D}/\omega = \pm 10 $ and (ii) a single peak located at $\mathcal{D}/ \omega = - 1$. Peaks (i) are not related to the phonon mode. They are located at $\mathcal{D} = \pm \omega_x$,  have smaller amplitudes and originate from the x-direction driving field~\cite{Core-Statis}.
Peak (ii), on the other hand, is induced by the spin-phonon coupling, reveals the phonon resonance $\mathcal{D} / \omega = -1$, and has larger amplitude. A further increase of the driving-field amplitude smears the three-peak picture out.

\begin{figure}[b]
\adjustimage{width=.5\textwidth,right}{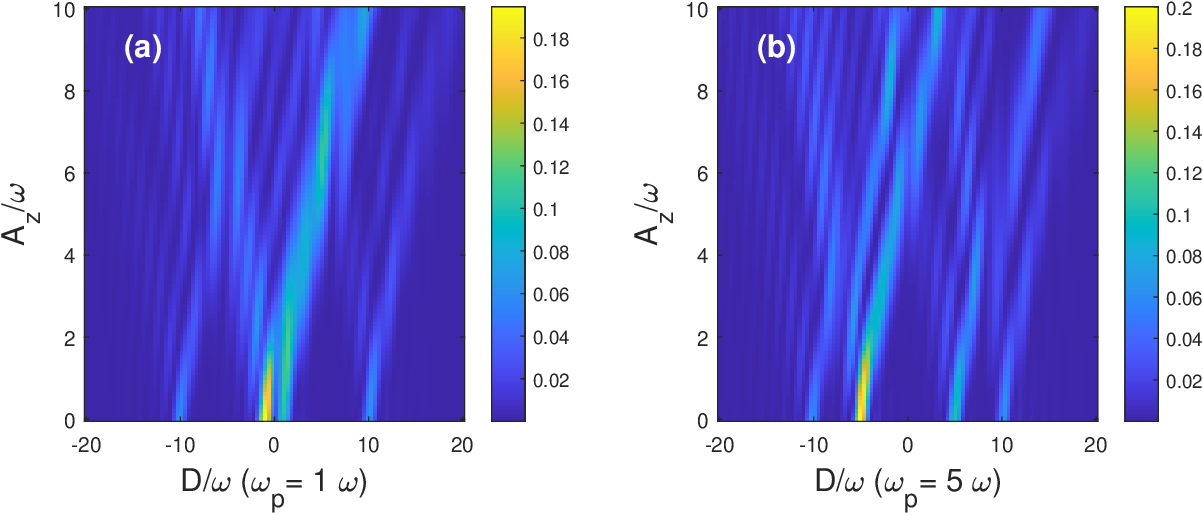}
\caption{ Panel (a): Contour plot of $P_{-1} ( t = 5/\omega)$ with the phonon initial displacement $f_{kn} = 1$. The remaining parameters are the same as in Fig.~\ref{Fig3}. Panel (b): Same parameters as in panel (a) but for  the phonon frequency  $\omega_p/\omega = 5 $.}
\label{Fig4}
\end{figure}

If off-diagonal phonon coupling and the $x$-direction driving field play similar role, then another symmetric peak at $\mathcal{D} /\omega = 1$ is expected to appear. However, only a single peak matching  the phonon frequency is found in Fig.~\ref{Fig3}. The absence of the second peak is caused by our choice of the initial condition:    displacements of the phonon coherent states  at $t = 0$ are set to zero, $f_{nk}(0)=0$.
For verification, Fig.~\ref{Fig4}(a) shows the contour plot evaluated for $f_{nk}(0)=1$. In this case, the amplitude of the peak at $\mathcal{D}/ \omega = 1$ is lower than that of the peak at $\mathcal{D}/ \omega = -1$. From the Fock-state expansion of the coherent state, $\ket{\alpha} = e^{-{|\alpha|}^2/2}\sum_{{n} = 0}^{\infty} {\alpha^n} \ket{{n}} /{\sqrt{{n}!}}$, it is obvious that the vacuum state takes up $60.65 \%$ of the initial Ansatz $\ket{\alpha = 1}$. Hence only 39.35 \% of the initial amplitude survives application of $\hat{b}_k$ (vibrational deactivation), while 100\% of the initial amplitude  survives application of $\hat{b}^{\dagger}_k$ (vibrational excitation). This leads to the unequal peak amplitudes at $\mathcal{D} / \omega = \pm 1$.
In Fig.~\ref{Fig4}(b), $\omega_p /\omega$ is increased to $5$ to avoid overlap with the phonon-induced peaks.
The peak shapes for $\mathcal{D}/\omega  = \pm 5 $ in Fig.~\ref{Fig4}(b) are more fragmented in comparison with those in Fig.~\ref{Fig3}.
This demonstrates that non-zero initial displacements  promote transitions to the Fock states with higher phonon numbers and lead to more complex dynamics of  $P_{-1} (t)$.
Figs.~\ref{Fig3} and \ref{Fig4} reveal also  that the phonon-induced peak heights exceed those induced by the driving fields, despite that $\eta_x = A_x$. This implies that the  tunneling effect induced by the off-diagonal phonon coupling is stronger than that due to the $x$-direction driving field in this anisotropic system.

\begin{figure}[h]
\vspace{-0.6cm}
\centerline{\includegraphics[width=92mm,scale=1]{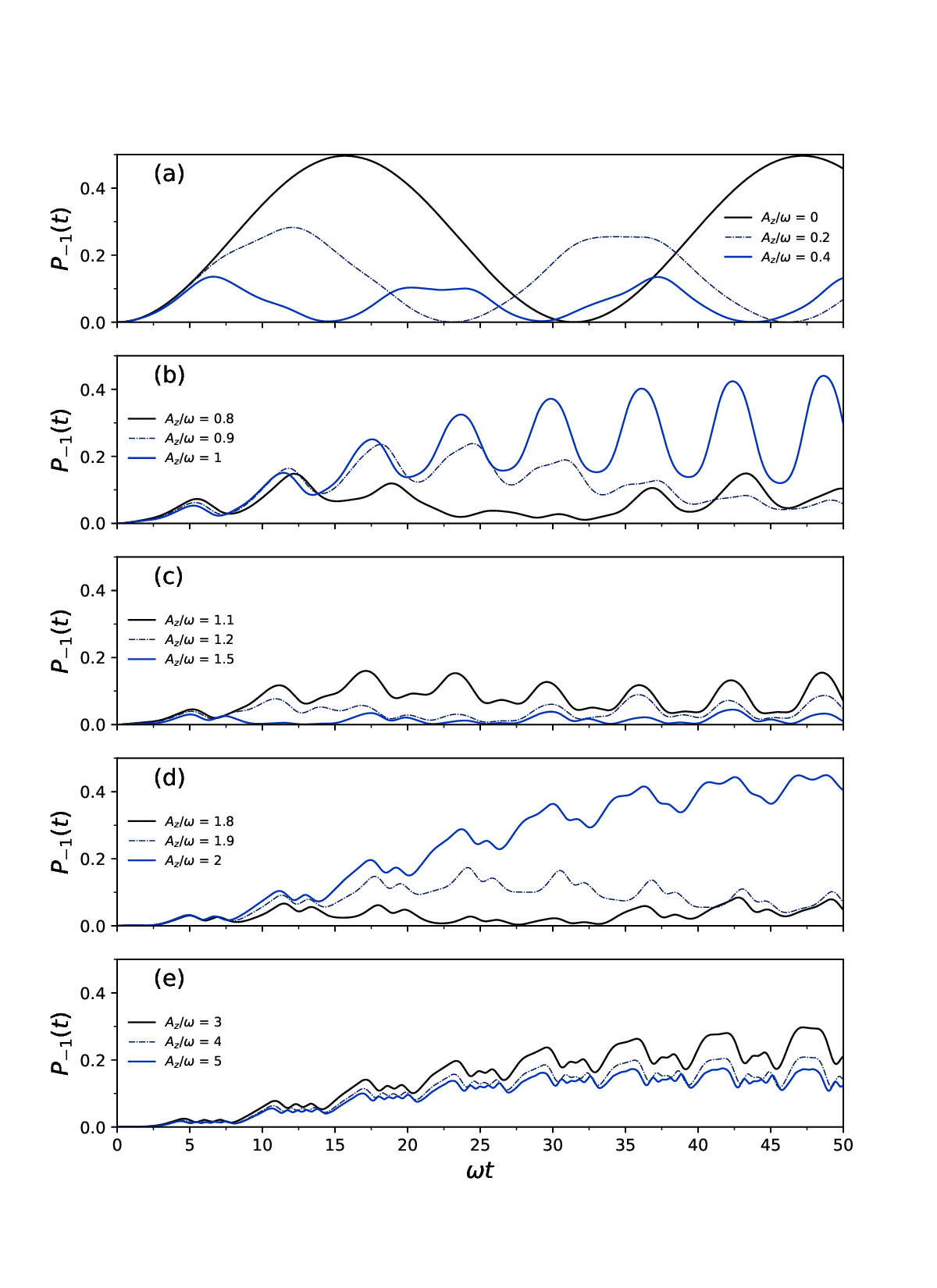}}
\vspace{-0.6cm}
\caption{$P_{-1} (t)$ for $\mathcal{D}$ and $A_z$ sampled from the brightest strip of the phonon-induced peak in the contour plot of Fig.~\ref{Fig3}, where $A_z$ and $\mathcal{D}$ obey the relation of $\mathcal{D} = -\omega+A_z$. Other parameters are $A_x/\omega = 0.1$, $\omega_{x}/\omega = 10$, $\omega_{p} /\omega= 1$, $\eta_{x}/\omega = 0.1$ and $\eta_{z}/\omega = 0$.  Panel (a): $A_z/\omega = 0, 0.2$ and $ 0.4$; Panel (b): $A_z/\omega = 0.8, 0.9$ and $1$; Panel (c): $A_z/\omega = 1.1, 1.2$ and $ 1.5$; Panel (d): $A_z/\omega = 1.8, 1.9$ and $2$; Panel (e): $A_z/\omega = 3, 4$ and$ 5$.}
\label{Fig5}
\end{figure}

In the contour plots, the peaks at $\mathcal{D}/\omega = - 1$ have high amplitudes in the segment from $A_z = 0$ to $A_z/\omega = 1.5$. These peaks reside on the line $\mathcal{D}/\omega = -1+A_z /\omega$.  Fig.~\ref{Fig5} shows populations $P_{-1} (t)$ evaluated up to  $\omega t =50$ for several representative  $(\mathcal{D}=-\omega +Az, \, A_z)$. The remaining parameters are the same as for Fig.~\ref{Fig3}. For better visibility, the curves are grouped into 5 panels, from (a) to (e).
In Fig.~\ref{Fig5}(a),  populations $P_{-1} (t)$ oscillate in a sinusoidal manner. As $A_z$ increases, both amplitudes and periods of the oscillations decrease. In Fig.~\ref{Fig5}(b), the behavior changes: populations start to exhibit an overall growth superimposed with oscillations of increasing amplitude and a period of $\sim 2\pi$ corresponding to the oscillation frequency of 1. In Fig.~\ref{Fig5}(c), the values of $P_{-1}(t)$ decrease significantly and beatings change their shapes.  In Fig.~\ref{Fig5}(d), the overall populations increase  as $A_z/\omega$ approaches $2$, and oscillatory features attain internal structure. Comparing the oscillation pattern at $A_z/\omega = 1$ at $A_z/\omega = 2$, for example, one recognizes that the  period of oscillations remains the same, while their amplitudes are significantly smaller at $A_z/\omega = 2$. In Fig.~\ref{Fig5}(e), as $A_z/\omega$ continues to increase,   $P_{-1} (t)$ show a gradual decrease  retaining the same shape. It is noteworthy that $P_{-1} (t)$ becomes very small for certain values of the parameters, for example, for $A_z/\omega=1.5$ and 1.8. This indicates that the tunneling effect due to the off-diagonal phonon coupling is almost absent. This can be attributed to CDT, where tuning of the driving frequency and other parameters strongly suppresses tunneling in quantum systems \cite{CDT, CDT1, CDT2, CDT3}.

Now we discuss in more detail transformations of the oscillatory patterns in Fig.~\ref{Fig5}.
If $A_z = 0$, then $\mathcal{D}/\omega= -1$ so that $|\mathcal{D}|$ is 10 times smaller than $\omega_x$. As a result, the $x$-direction driving field has little influence on the 3-LZM dynamics. Hence the tunneling is caused  mostly by the off-diagonal phonon coupling, which induces Rabi oscillations in $P_{-1} (t)$ shown in Fig.~\ref{Fig5}(a).
For $A_z/\omega = 1$, $P_{-1} (t)$ increases with time and exhibits oscillations with a period $2\pi / \omega$, consistent with the frequency of the $z$-direction driving field (Fig.~\ref{Fig5}(b)). For $A_z/\omega = 2$,  amplitudes of the fast oscillations are significantly reduced, as displayed in Fig.~\ref{Fig5}(d). A possible explanation is as follows. If one approximates the periodic driving field at the avoided crossings by the linear driving field, the scanning velocity of the linear driving will be higher if the amplitude of the periodic field is larger. High scanning velocities at the avoided crossing  mitigate the effect of the $z$-direction driving and decease the high-frequency oscillations. This also enlarges the gap between the energy levels at the avoided crossing,  weakens the tunneling effect and thereby decreases the  $P_{-1} (t)$, as seen in Fig.~\ref{Fig5}(e). The shapes of the oscillatory features in Figs.~\ref{Fig5}(d) and (e) exhibit internal structure, which is caused by the strong external driving and coupling to the phonon modes. Similar phenomena were demonstrated in dissipative dynamics of nonadiabatic systems \cite{Gelin09}.

\begin{figure}[b]
\centerline{\includegraphics[width=90mm,scale=1]{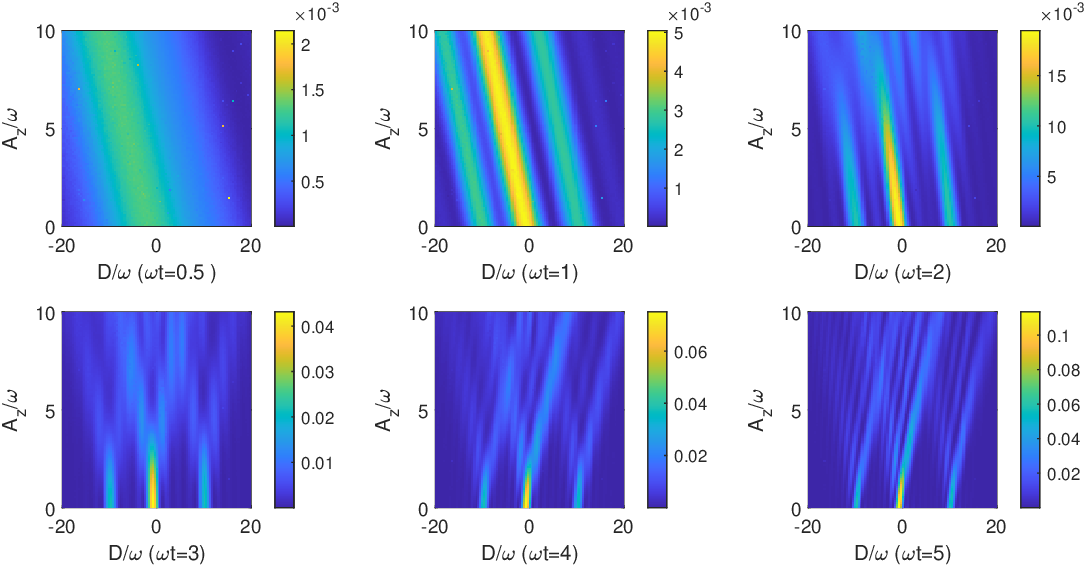}}
\caption{Time evolution of the contour plot of $P_{-1} (t)$ from ${\omega t = 0}$ to ${\omega t = 5}$. The parameters are identical to those in Fig.~\ref{Fig3}. A video that captures the time evolution can be found in supplementary materials.}
\label{Fig6}
\end{figure}

\

The contour plot in Fig.~\ref{Fig3} corresponds to a specific time $\omega t = 5$. It is  elucidating, however, to follow the time evolution of the contour plots.  This is done in Fig.~\ref{Fig6} which displays six snapshots of the contour plot,  from $\omega t = 0$ to $\omega t = 5$. At $\omega t = 0.5$, the features are smeared out and the overall transition probability is small, because the system does not have enough time for tunneling. As $t$ increases, the $A_z-\mathcal{D}$ patterns  gradually become sharper and narrower, and transition probabilities increase. In addition,  as the system evolves, the stripes in the plots rotate clockwise. These observations are caused by  Rabi cycling. For nearly resonance points in the contour plot, period and amplitudes of Rabi oscillations are large. Because of the large period, populations of the nearly resonance points rise slowly. On the other hand, population of the off-resonance points rise much faster owing to their smaller periods. Hence they form broad and blurry structures at short $t$. At $t$ growths further, populations of the off-resonance points remain small while populations of the nearly resonance points continue growing and eventually become the major contributors to the striped structures of the contour plot. At even  longer times, the resonance pattern is predicted to narrow and eventually evolve into separate lines revealing exact resonances. To better visualize the time evolution of the contour plot, the reader is referred to the supplementary video.

\subsection{Multi-mode phonon coupling with periodic driving field}

In Secs. III A and B, we considered a single-mode bath. In this subsection we extend our discussion to the dynamics of the 3-LZM coupled to the multi-mode bath.

We consider the super-Ohmic bath with the spectral density of Eq.~(\ref{EQ3}) with $s = 3$ which is coupled to the 3-LZM in the $z$-direction. We set $\mathcal{D} /\omega= \omega_x /\omega = 10$, assuming that the 3-LZM is in resonance with the $x$-direction driving field of the amplitude  $A_x /\omega= 0.05$. The $z$-direction driving field of the amplitude $A_z/\omega = 0.5$ is also included in the calculations. The cut-off frequency $\omega_c/\omega$ is fixed at  $0.5$.  Following Section II A, the continuum spectrum is discretized into 20 phonon modes ($N_b = 20$) with $\omega_k \in [0, \omega_m]$. In this case,  $\omega_c$, $\omega_m$ and $N_b$ are large enough, their specific values have no effect on the evaluated observables, and multiplicity of $M = 16$ ensures converged results.

\begin{figure}[bt]
\centerline{\includegraphics[width=80mm,scale=0.8]{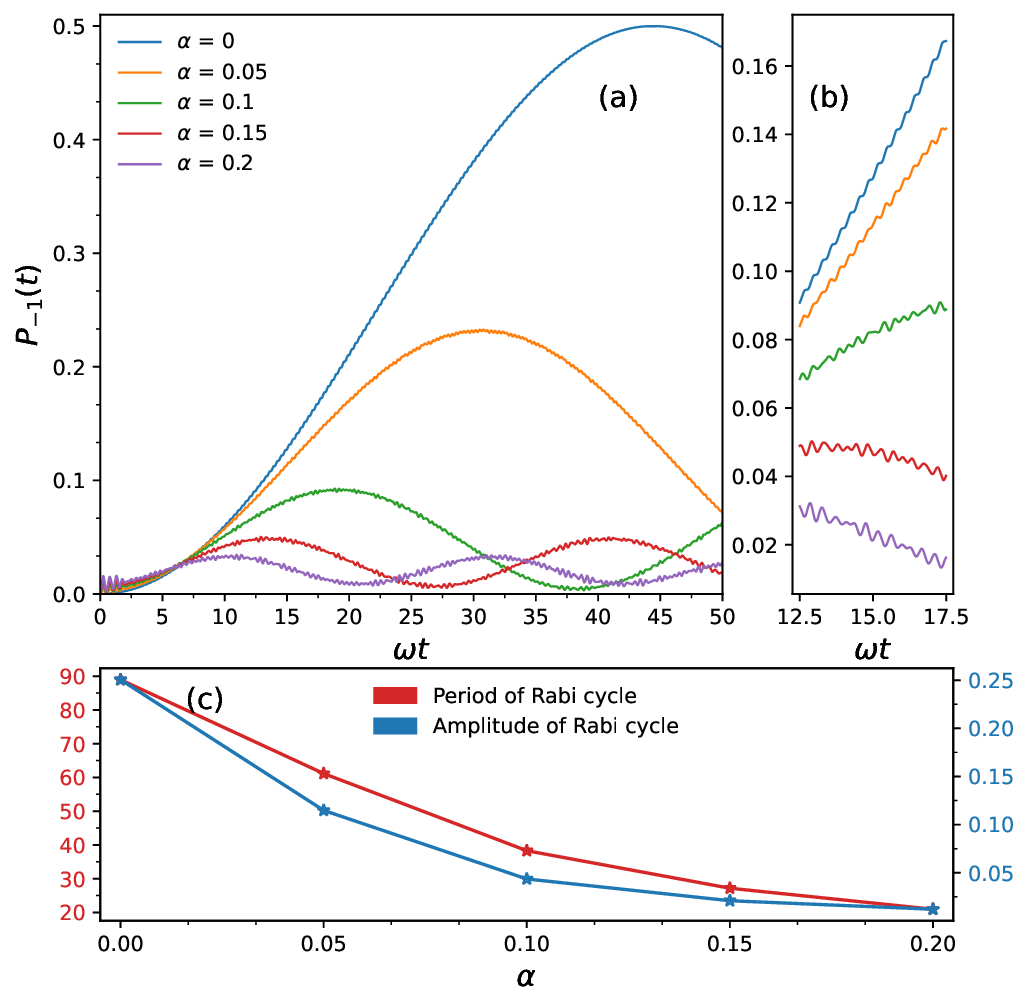}}
\caption{Panel (a):  Time evolution of the population $P_{-1} (t)$ from $\omega t = 0$ to $\omega t = 50$ at five values of $\alpha$ indicated in the legend. Other parameters are as follows.  $\hat{H}_{\rm 3-LZM}$: $\mathcal{D}/\omega = 10$, $A_z/\omega= 0.5$, $A_x/\omega = 0.05$, and $\omega_x/\omega = 10 $. $\hat{H}_{\rm sp}$: $s = 3$, $\omega_c /\omega= 0.5$, $\omega_m/\omega = 6$, $N_b = 20$. Panel (b): amplified portion of the figure in panel (a) from $\omega t = 12.5$ to $\omega t = 17.5$. Panel (c): The periods (red) and the amplitudes (blue) of the Rabi oscillations obtained via sinusoidal fitting of the curves in panel (a)  plotted as a function of $\alpha$.}
\label{Fig7}
\end{figure}

In Fig.~\ref{Fig7}(a), the time evolutions of $P_{-1} (t)$ are plotted  for five values of the bath coupling strength $\alpha$ ranging from 0 to 0.2. In the absence of the bath ($\alpha = 0$), $P_{-1} (t)$ exhibits Rabi oscillations (cf.  Fig.~\ref{Fig5}(a)). As coupling to the bath becomes stronger, Rabi frequencies  increase but the amplitudes of Rabi oscillations decrease. Interestingly, growing  $A_x$ also increase frequencies and decrease amplitudes of the  $P_{-1} (t)$ oscillations in the bath-free case, causing, simultaneously, changes in shapes of the oscillatory features (cf. Fig.~\ref{Fig5}(a)). This observation may help to distinguish between the $x$-field driving and the bath effects.

In Fig.~\ref{Fig7}(b), we zoom in a portion of Fig.~\ref{Fig7}(a) from $\omega t = 12.5$ to $\omega t = 17.5$. Apart from Rabi beatings, high-frequency low-amplitude oscillations in $P_{-1} (t)$ are clearly visible in Fig.~\ref{Fig7}(b). Similar oscillations were found in the
dissipative dynamics of a Rabi dimer with periodic harmonic driving \cite{QED_Huang}.
At $\alpha = 0$, the oscillations are quite regular, with a period around 0.3/$\omega $. This indicates that the high-frequency oscillations are likely caused by the $x$-direction driving field which has a period of ${2\pi}/{\omega_x}=0.63$. However,  such high-amplitude oscillations cannot be found in Fig.~\ref{Fig5}(a), where $A_z/\omega = 0$. This indicates that the oscillations are, most likely, caused by the beatings of the driving fields in $x$ and $z$ directions. Note also that the difference between the corresponding curves in Fig.~\ref{Fig5}(a) and Fig.~\ref{Fig7}(b) occurs despite the $x$-direction phonon coupling and the $x$-direction driving field may produce peaks with similar shapes in the contour plots.
As $\alpha$ increases, the high-frequency low-amplitude oscillations become less regular but do not disappear. This is a signature of the combined effect of the $x$-direction driving field and the phonon coupling.

To quantify the bath-induced  quenching, the periods and amplitudes of the $P_{-1} (t)$ oscillations obtained via sinusoidal fitting are plotted in Fig.~\ref{Fig7}(c) as a function of the system-bath coupling $\alpha$. It is found that the amplitude of Rabi oscillations drops more than twice for the couplings varying from $\alpha = 0$ to $\alpha = 0.05$. At larger $\alpha$, the rate of decrease of the amplitude slows down. A similar mitigation of the bath impact on the period of Rabi oscillations is also observed.

\section{Conclusion}

We have developed the methodology for numerically accurate simulation of the dynamics  3-LZMs driven by external fields of any shape and coupled to bosonic baths with arbitrary spectral densities.
The methodology is based on the multi-D$_2$ Davydov Ansatz, in which bosonic environment is accounted for through an expansion of the total (system + bath) many-body wave function in multiple coherent states per bath mode. The so-constructed wave function yields exact solution of the multidimensional time-dependent Schr\"odiger equation if multiplicity of the  Ansatz is sufficiently high.

We applied the multi-D$_2$ methodology for performing numerically accurate simulations of the 3-LZM coupled to a vibrational reservoir. We investigated how the combined effect of anisotropy,  external (linear and harmonic) driving fields in $x$ and $z$ directions, and phonon modes modify the  3-LZM  population evolutions. It is found that phonon modes give rise to a plethora of different effects, from opening up additional phonon-assisted channels for LZ transitions in the case of a single phonon mode to reducing coherent effects and initiating nontrivial modifications of the 3-LZM dynamics in the case of a  multidimensional phonon bath. While it is no surprise that dynamic behaviors of the chosen anisotropic 3-LZM are extraordinarily rich and complex, we have tried to establish, whenever possible, how signatures of individual contributions (e.g., external driving or phonon coupling) can be identified in the ensuing population evolutions.

The numerically accurate Davydov-Ansatz methodology developed in the present work is exceptionally powerful.
On the one hand, it can be used for benchmarking various approximate (e.g., based on master equations) methods of the description of driven 3-LZMs coupled to phonon modes/baths. On the other hand, it is applicable for arbitrary time-dependent driving fields as well as for boson baths with any spectral densities and, if necessary, at final temperatures. This may facilitate future utilization of the developed simulation machinery for studying specific experimental realizations of 3-LZM systems  and using external fields for optimizing and controlling LZ transitions in such systems. If necessary, the methodology can be straightforwardly generalized towards N-LZMs.

\section*{Authors' contributions}
L.Z. and L.W. contributed equally to this work.

\section*{Acknowledgments}
The authors would also thank Dr. Kewei Sun and Xia Yang for useful discussion.
Support from Nanyang Technological University ``URECA" Undergraduate Research Programme and
the Singapore Ministry of Education Academic Research
Fund Tier 1 (Grant No. RG87/20) is gratefully acknowledged.
M. F. G. acknowledges support from the Hangzhou Dianzi University
through startup funding.
\section*{Author Declarations}
\subsection*{Conflict of Interest}
The authors have no conflicts to disclose.

\section*{Data Availability}
The data that support the findings of this study are available from the corresponding author upon reasonable request.

\appendix
\onecolumngrid
\section{The equation of motions for the multi-D$_2$ Ansatz}\label{Appendix A}

The Lagrangian in Eq.~(\ref{EQ10}) can be simplified into the following form due to the normalization of the Ansatz state:
\begin{eqnarray}\label{EQA1}
L&=&i\langle {\rm D}_{2}^{M}(t)|\frac{\overrightarrow{\partial}}{\partial t}|{\rm D}_{2}^{M}(t)\rangle -\langle{\rm D}_{2}^{M}(t)|\hat{H}_{\rm sys }+\hat{H}_{\rm ph}|{\rm D}_{2}^{M}(t)\rangle = L_d - L_H
\end{eqnarray}

with:
\begin{eqnarray}\label{EQA2}
L_d &=&i\langle {\rm D}_{2}^{M}(t)|\frac{\overrightarrow{\partial}}{\partial t}|{\rm D}_{2}^{M}(t)\rangle \nonumber\\
& = &i\sum_{m, n=1}^{M}\left(A_{m}^{*}\dot{A}_{n}+B_{m}^{*}\dot{B}_{n}+C_{m}^{*}\dot{C}_{n}\right)S_{mn}\nonumber\\
&&+ i\sum_{m, n=1}^{M}\left(A_{m}^{*}A_{n}+B_{m}^{*}B_{n}+C_{m}^{*}C_{n}\right)\left(f_{mk}^{*}\dot{f}_{nk}-\frac{1}{2}\dot{f}_{nk}f_{nk}^{*}-\frac{1}{2}f_{nk}\dot{f}_{nk}^{*}\right)S_{mn}
\end{eqnarray}

and
\begin{eqnarray}\label{EQA3}
L_H&=&\langle{\rm D}_{2}^{M}(t)|\hat{H}_{\rm sys }+\hat{H}_{\rm ph}|{\rm D}_{2}^{M}(t)\rangle \nonumber\\
& = & \sum_{m, n=1}^{M} \big[ \Omega_z\left(A_{m}^{*}A_{n}-C_{m}^{*}C_{n}\right)+\Omega_x\left(A_{m}^{*}B_{n}+B_{m}^{*}A_{n}+B_{m}^{*}C_{n}+C_{m}^{*}B_{n}\right) \big]S_{mn}\nonumber\\
&&+ \sum_{m, n=1}^{M}\big[D\left(\frac{1}{3}A_{m}^{*}A_{n}-\frac{2}{3}C_{m}^{*}C_{n}+\frac{1}{3}C_{m}^{*}C_{n}\right)+\sum_k \omega_k \left(A_{m}^{*}A_{n}+B_{m}^{*}B_{n}+C_{m}^{*}C_{n}\right)f_{mk}^{*}f_{nk}\big]S_{mn}\nonumber\\
&&+ \sum_{m, n=1}^{M}\sum_k\big[\eta_k^z\left(A_{m}^{*}A_{n}-C_{m}^{*}C_{n}\right)+\frac{\eta_k^x}{\sqrt{2}}\left(A_{m}^{*}B_{n}+B_{m}^{*}A_{n}+B_{m}^{*}C_{n}+C_{m}^{*}B_{n}\right)\big]\left(f_{mk}^{*}+f_{nk}\right)S_{mn}
\end{eqnarray}

Following the Dirac-Frenkel time dependent variational principle, a set of EOMs can be solved by substituting $u_n$ in Eq.~(\ref{EQ10}) with $A_m$:
\begin{eqnarray}\label{EQA4}
&~& \sum_{n=1}^{M}\left[\mathrm{i}\dot{A}_{n}-\frac{\mathrm{i}}{2}A_{n}\sum_{k}\left(f_{nk}^{*}\dot{f}_{nk}+f_{nk}\dot{f}_{nk}^{*}-2f_{mk}^{*}\dot{f}_{mk}\right)\right]S_{mn}\nonumber\\
&= & \sum_{n=1}^{M}\left[ A_{n}\left( \Omega_z(t)+\frac{D}{3}+\sum_{k}[\omega_{k}f_{mk}^{*}f_{nk}+\eta_{k}^{z}\left(f_{mk}^{*}+f_{nk}\right)]\right )+ B_{n}\left(\Omega_x(t)+\sum_{k}\frac{\eta_{k}^{x}}{\sqrt{2}}\left(f_{mk}^{*}+f_{nk}\right)\right)\right]S_{mn}
\end{eqnarray}

Similarly with $B_m$:

\begin{eqnarray}\label{EQA5}
&~& \sum_{n}\left[\mathrm{i}\dot{B}_{n}-\frac{\mathrm{i}}{2}B_{n}\sum_{k}\left(f_{nk}^{*}\dot{f}_{nk}+f_{nk}\dot{f}_{nk}^{*}-2f_{mk}^{*}\dot{f}_{nk}\right)\right]S_{mn}\nonumber\\
&= & \sum_{n=1}^{M}\left[B_{n}\left(-\frac{2}{3}D+\sum_{k}\omega_{k}f_{mk}^{*}f_{nk}\right)+\left(A_{n}+C_{n}\right)\left( \Omega_x(t)+\sum_{k}\frac{\eta_{k}^x}{\sqrt{2}}\left(f_{mk}^{*}+f_{nk}\right)\right)\right]S_{mn}
\end{eqnarray}

and with $C_m$:

\begin{eqnarray}\label{EQA6}
&~& \sum_{n=1}^{M}\left[\mathrm{i}\dot{C}_{n}-\frac{\mathrm{i}}{2}C_{n}\sum_{k}\left(f_{nk}^{*}\dot{f}_{nk}+f_{nk}\dot{f}_{nk}^{*}-2f_{mk}^{*}\dot{f}_{mk}\right)\right]S_{mn}\nonumber\\
&= & \sum_{n=1}^{M}\left[ C_{n}\left( -\Omega_z(t)+\frac{D}{3}+\sum_{k}[\omega_{k}f_{mk}^{*}f_{nk}+\eta_{k}^{z}\left(f_{mk}^{*}+f_{nk}\right)]\right )+ B_{n}\left(\Omega_x(t)+\sum_{k}\frac{\eta_{k}^{x}}{\sqrt{2}}\left(f_{mk}^{*}+f_{nk}\right)\right)\right]S_{mn}
\end{eqnarray}

Lastly with $f_{mk}$:

\begin{eqnarray}\label{EQA7}
&~& \mathrm{i}\sum_{n=1}^{M}\left[\left(A_{m}^{*}\dot{A}_{n}+B_{m}^{*}\dot{B}_{n}+C_{m}^{*}\dot{C}_{n}\right)f_{nk}\right]S_{mn}\nonumber\\
&&+ \mathrm{i}\sum_{n=1}^{M}\left[\left(A_{m}^{*}{A}_{n}+B_{m}^{*}{B}_{n}+C_{m}^{*}{C}_{n}\right)\left(\dot{f}_{nk}-\frac{1}{2}f_{nk}\sum_{k^\prime}\left[\left(f_{nk^\prime}^{*}-2f_{mk^\prime}^{*}\right)\dot{f}_{nk^\prime}+f_{nk^\prime}\dot{f}_{nk^\prime}^{*}\right]\right)\right]S_{mn}\nonumber\\
&= & \sum_{n=1}^{M}\left(A_{m}^{*}A_{n}-C_{m}^{*}C_{n}\right)\left[\left(\Omega_z(t)+\sum_{k^\prime}\eta_{k^\prime}^{z}\left(f_{mk^\prime}^{*}+f_{nk^\prime}\right)\right)f_{nk}+\eta_{k}^{z}\right]S_{mn}\nonumber\\
&& +\sum_{n=1}^{M}\left(A_{m}^{*}A_{n}+B_{m}^{*}B_{n}+C_{m}^{*}C_{n}\right)\left[f_{nk}\left(\sum_{k^\prime}\omega_{k^\prime}f_{mk^\prime}^{*}f_{nk^\prime}\right)+\omega_{k}f_{nk}\right]S_{mn}\nonumber\\
&& +\sum_{n=1}^{M}\left(A_{m}^{*}B_{n}+B_{m}^{*}A_{n}+B_{m}^{*}C_{n}+C_{m}^{*}B_{n}\right)\left[\left(\Omega_x(t)+\sum_{k^\prime}\frac{\eta_{k^\prime}^{x}}{\sqrt{2}}\left(f_{mk^\prime}^{*}+f_{nk^\prime}\right)\right)f_{nk}+\frac{\eta_{k}^{x}}{\sqrt{2}}\right]S_{mn}\nonumber\\
&& +\sum_{n=1}^{M}\left(A_{m}^{*}A_{n}-2B_{m}^{*}B_{n}+C_{m}^{*}C_{n}\right)\frac{D}{3}f_{nk}S_{mn}
\end{eqnarray}

\twocolumngrid

\begin{thebibliography}{0}
\expandafter\ifx\csname natexlab\endcsname\relax\def\natexlab#1{#1}\fi
\expandafter\ifx\csname bibnamefont\endcsname\relax
  \def\bibnamefont#1{#1}\fi
\expandafter\ifx\csname bibfnamefont\endcsname\relax
  \def\bibfnamefont#1{#1}\fi
\expandafter\ifx\csname citenamefont\endcsname\relax
  \def\citenamefont#1{#1}\fi
\expandafter\ifx\csname url\endcsname\relax
  \def\url#1{\texttt{#1}}\fi
\expandafter\ifx\csname urlprefix\endcsname\relax\def\urlprefix{URL }\fi
\providecommand{\bibinfo}[2]{#2}
\providecommand{\eprint}[2][]{\url{#2}}

\end{thebibliography}


\begin{thebibliography}{999}
\bibitem{Landau}L. D. Landau, Zur Theorie der Energie\"ubertragung. II, Phys. Z. Soviet Union {\bf 2}, pp. 46-51 (1932)

\bibitem{Zener} C. Zener. Non-adiabatic crossing of energy levels. Proc. R. Soc. Lond. Ser. A Math. Phys. Eng. Sci. {\bf 137}, 696-702 (1932)

\bibitem{m_ph1}A. Niranjan, W. Li, and R. Nath, Landau-Zener transitions and adiabatic impulse approximation in an array of two Rydberg atoms with time-dependent detuning, Phys. Rev. A {\bf 101},063415 (2020)


\bibitem{m_ph2}S. S. Zhang and W. Gao and H. Cheng and L. You and H. P. Liu, Symmetry-Breaking Assisted Landau-Zener Transitions in Rydberg Atoms, Phys. Rev. Lett. {\bf 120}, 063203 (2018)


\bibitem{LZ-cavity1}N. A. Sinitsyn and F. Li, Solvable multistate model of Landau-Zener transitions in cavity QED, Phys. Rev. A {\bf 93}, 063859 (2016)


\bibitem{chem_physics1} L. Zhu, A. Widom and P. M. Champion, A multidimensional Landau-Zener description of chemical reaction dynamics and vibrational coherence, J. Chem. Phys. {\bf 107}, 2859 (1997)


\bibitem{QIP1}Cao, G., Li, HO., Tu, T. et al, Ultrafast universal quantum control of a quantum-dot charge qubit using Landau–Zener–St\"uckelberg interference, Nat Commun {\bf 4}, 1401 (2013)


\bibitem{QIP2}Matityahu, S., Schmidt, H., Bilmes, A. et al. Dynamical decoupling of quantum two-level systems by coherent multiple Landau–Zener transitions, npj Quantum Inf {\bf 5}, 114 (2019)



\bibitem{nanomagnet} W. Wernsdorfer,  R. Sessoli,  A. Caneschi,  D. Gatteschi,  A. Cornia,  Nonadiabatic Landau-Zener tunneling in $Fe_8$ molecular nanomagnets, Europhys. Lett. {\bf 50}, 552 (2000)


\bibitem{BEC}Abraham J. Olson, Su-Ju Wang, Robert J. Niffenegger, Chuan-Hsun Li, Chris H. Greene, and Yong P. Chen, Tunable Landau-Zener transitions in a spin-orbit-coupled Bose-Einstein condensate, Phys. Rev. A {\bf 90}, 013616 (2014)


\bibitem{quenching} L. Arceci, S. Barbarino, R. Fazio, and G. E. Santoro, Dissipative Landau-Zener problem and thermally assisted Quantum Annealing, Phys. Rev. B {\bf 96}, 054301 (2017)


\bibitem{NV1}J. Zhou, P. Huang, Q. Zhang et al, Observation of Time-Domain Rabi Oscillations in the Landau-Zener Regime with a Single Electronic Spin, Phys. Rev. Lett. {\bf 112}, 010503 (2014)


\bibitem{NV2}P. Huang, J. Zhou, F. Fang, X. Kong, X. Xu, C. Ju, J.Du, Landau-Zener-St\"uckelberg Interferometry of a Single Electronic Spin in a Noisy Environment, Phys. Rev. X {\bf 1}, 011003 (2011)


\bibitem{QDot1} X. Mi, S. Kohler, and J. R. Petta, Landau-Zener interferometry of valley-orbit states in Si/SiGe double quantum dots,
Phys. Rev. B {\bf 98}, 161404(R) (2018)


\bibitem{QDot2}P. Nalbach, J. Kn\"orzer, and S. Ludwig, Nonequilibrium Landau-Zener-Stueckelberg spectroscopy in a double quantum dot,
Phys. Rev. B {\bf 87}, 165425 (2013)


\bibitem{LZ_review}O. V. Ivakhenko, S. N. Shevchenko, F. Nori, Nonadiabatic Landau–Zener–St\"{u}ckelberg–Majorana transitions, dynamics, and interference , Phys. Rep. {995}, pp. 1-89 (2023)


\bibitem{LZ-multi1}V. Chernyak, N. Sinitsyn, C. Sun, Multitime Landau–Zener model: classification of solvable Hamiltonians, J. Phys. A: Math. Theor. {\bf 53} ,185203 (2020)


\bibitem{Volodya21}R. K. Malla,  V. Y. Chernyak,  N. A. Sinitsyn. Nonadiabatic transitions in Landau-Zener grids: integrability and semiclassical theory
 Phys. Rev. B {\bf 103}, 144301 (2021).



\bibitem{Andrey}A. R. Kolovsky and D. N. Maksimov, Mott-insulator state of cold atoms in tilted optical lattices: Doublon dynamics and multilevel Landau-Zener tunneling, Phys. Rev. A {\bf 94}, 043630 (2016)


\bibitem{band}Y. B. Band, Y. Avishai, Three-level Landau-Zener dynamics, Phys. Rev. A{\bf 99}, 032112 (2019)


\bibitem{LZ-BEC-well_trap}E. M. Graefe, H. J. Korsch, and D. Witthaut, Mean-field dynamics of a Bose-Einstein condensate in a time-dependent triple-well trap: Nonlinear eigenstates, Landau-Zener models, and stimulated Raman adiabatic passage, Phys. Rev. A {\bf 73}, 013617 (2006)


\bibitem{at_gas}L. Cornelius Fai, M. Tchoffo, and M. Nana Jipdi, Landau-Zener scenario in a trapped atomic gas: multi-level multi-particle model, Eur. Phys. J. B {\bf 88}, 181 (2015)


\bibitem{TQD1}G. Granger, L. Gaudreau, A. Kam et al, Three-dimensional transport diagram of a triple quantum dot, Phys. Rev. B {\bf 82}, 075304 (2010)


\bibitem{TQD2}D. Schr\"{o}er, A. D. Greentree, L. Gaudreau, K. Eberl, L. C. L. Hollenberg, J. P. Kotthaus, and S. Ludwig, Electrostatically defined serial triple quantum dot charged with few electrons, Phys. Rev. B {\bf 76}, 075306 (2007)



\bibitem{D2-dw3}Z. Huang and Y. Zhao, Dynamics of dissipative Landau-Zener transitions, Phys. Rev. A {\bf 97}, 013803 (2018)


\bibitem{LZ-periodic2}F. Zheng, Y. Shen, K. Sun, Y. Zhao, Photon-assisted Landau–Zener transitions in a periodically driven Rabi dimer coupled to a dissipative mode, J. Chem. Phys. {\bf 154}, 044102 (2021)


\bibitem{LZ-entanglement1}C. Quintana, K. Petersson, L. McFaul, S. Srinivasaan, A. Houck, J. Petta, Cavity-mediated entanglement generation via Landau-Zener interferometry, Phys. Rev. Lett. {\bf 110}, 173603 (2013)


\bibitem{LZ-entanglement2}M. Wubs, S. Kohler, and P. H\"anggi, Entanglement creation in circuit QED via Landau-Zener sweeps, Physica E Low Dimens. Syst. Nanostruct. {\bf 40}, 187-197 (2007)



\bibitem{bath1}K. Saito, M. Wubs, S. Kohler, P. H\"anggi, Y. Kayanuma. Quantum state preparation in circuit QED
via Landau-Zener tunneling. Europhys. Lett. {\bf 76}, 22-28 (2006)


\bibitem{bath2} M. Wubs, K. Saito, S. Kohler, P. H\"anggi, and Y. Kayanuma, Gauging a Quantum Heat Bath with Dissipative Landau-Zener Transitions, Phys. Rev. Lett.  {\bf 97}, 200404 (2006)


\bibitem{YZ19} M. Werther,  F. Grossmann, Z. Huang,  Y. Zhao. Davydov-Ansatz for Landau-Zener-Stueckelberg-Majorana transitions in an environment: Tuning the survival probability via number state excitation. J. Chem. Phys. {\bf 150}, 234109 (2019)


\bibitem{Thorwart09} P. Nalbach, M. Thorwart. Landau-Zener Transitions in a Dissipative Environment: Numerically Exact Results, Phys. Rev. Lett. {\bf 103}, 220401 (2009)


\bibitem{Thorwart15}S. Javanbakht,  P. Nalbach, M. Thorwart. Dissipative Landau-Zener quantum dynamics with transversal and longitudinal noise, Phys. Rev.  A {\bf 91}, 052103 (2015)



\bibitem{dibatic_basis}K. Saito, M. Wubs, S. Kohler, Y. Kayanuma, P. H\"anggi, Dissipative Landau-Zener transitions of a qubit: Bath-specific and universal behavior, Phys. Rev. B {\bf 75}, 214308 (2007)




\bibitem{Core-kiselev}M. N. Kiselev, K. Kikoin, and M. B. Kenmoe, SU(3) Landau-Zener interferometry, Euro. Phys. Lett. {\bf 104}, 57004 (2013)


\bibitem{extention1}J. Lin, N. A. Sinitsyn, Exact transition probabilities in the three-state Landau–Zener–Coulomb model, J. Phys. A: Math. Theor. {\bf 47}, 015301 (2014)


\bibitem{extention2}C. E. Carroll, F. T. Hioe, Generalization of Landau-Zener calculation to three levels, J. Phys. A: Math. Gen. {\bf 19}, 1151-1161 (1986)


\bibitem{NV_review1}A. Gali, Ab initio theory of the nitrogen-vacancy center in diamond, Nanophotonics {\bf 8}, https://doi.org/10.1515/nanoph-2019-0154 (2019)


\bibitem{NV-computing1}M. V. Gurudev Dutt, L. Childress, L. Jiang et al, Quantum register based on individual electronic and nuclear spin qubits in diamond, Science {\bf 316},
1312-1316 (2007)


\bibitem{NV-computing2}S. Sangtawesin, C. McLellan, B. Myers, A. Jayich, D. Awschalom, J. Petta, Hyperfine-enhanced gyromagnetic ratio of a nuclear spin in diamond, New J. Phys. {\bf 18}, 083016 (2016)


\bibitem{NV-computing3}P. C. Maurer, G. Kucsko, C. Latta et al, Room-temperature quantum bit memory exceeding one second, Science {\bf 336}, 1283-1286 (2012)


\bibitem{NV-stress} M. Doherty, V. Struzhkin, D. Simpson et al, Electronic Properties and Metrology Applications of the Diamond $NV^{-}$ Center under Pressure, Phys. Rev. Lett. {\bf 112}, 047601 (2014)


\bibitem{TQD-automata}C. Lent, D. Tougaw, W. Porod, G. Bernstein, Quantum cellular automata, Nanotechnology {\bf 4}, 49 (1993)


\bibitem{TQD-rectifier}M. Stopa, Rectifying Behavior in Coulomb Blockades: Charging Rectifiers, Phys. Rev. Lett. {\bf 88}, pp.4 (2002)


\bibitem{Ashhab16}
S. Ashhab. Landau-Zener transitions in an open multilevel quantum system,
Phys. Rev. A  {\bf 94}, 042109 (2016)

\bibitem{Militello19a}B. Militello, Three-state Landau-Zener model in the presence of dissipation, Phys. Rev. A {\bf 99}, 033415 (2019)



\bibitem{Militello19b}B. Militello,
Detuning-induced robustness of a three-state Landau-Zener model against dissipation
Phys. Rev. A {\bf 99}, 063412 (2019)

\bibitem{Militello19c}B. Militello,  N. V. Vitanov,
Master-equation approach to the three-state open Majorana model, Phys. Rev.  A {\bf 100}, 053407 (2019)


\bibitem{D2-review}Y. Zhao, K. Sun, L. Chen, M. Gelin, The hierarchy of Davydov's Ans\"atze and its applications, Wiley Interdiscip. Rev. Comput. Mol. Sci. {\bf 12}, https://doi.org/10.1002/wcms.1589 (2022)


\bibitem{D2-HTC}L. Chen, and Y. Zhao, Finite temperature dynamics of a Holstein polaron: The thermo-field dynamics approach, J. Chem. Phys. {\bf 147}, 214102 (2017)

\bibitem{D2-TC}K.Sun, C. Dou, M. F. Gelin, Y. Zhao,  Dynamics of disordered Tavis-Cummings and Holstein-Tavis-Cummings models, J. Chem. Phys. {\bf 156}, 024102 (2022)



\bibitem{D2-SBM1}L. Wang, F. Zheng, J. Wang, F. Grossmann, Y. Zhao, Schr\"{o}dinger-Cat States in Landau-Zener-St\"{u}ckelberg-Majorana Interferometry: A Multiple Davydov Ansatz Approach, J. Phys. Chem. B {\bf 125}, 3184-3196 (2021)


\bibitem{D2-SBM2}L. Wang, L. Chen, N. Zhou, Y. Zhao,  Variational dynamics of the sub-Ohmic spin-boson model on the basis of multiple Davydov D1 states, J. Chem. Phys. {\bf 144}, 024101 (2016)




\bibitem{D2-SF}K.Sun, M. F. Gelin, and Y. Zhao, Accurate Simulation of Spectroscopic Signatures of Cavity-Assisted, Conical-Intersection-Controlled Singlet Fission Processes, J. Phys. Chem. Lett. {\bf 13}, 4280-4288 (2022).



\bibitem{Core-spec_den}M.L.Goldman, M. Doherty, A. Sipahigil et al,  State-selective intersystem crossing in nitrogen-vacancy centers, Phys. Rev. B Condens. Matter {\bf 91}, 165201 (2015)



\bibitem{WFCZ} L.~Wang, Y.~Fujihashi, L.~Chen,and Y.~Zhao, Finite-temperature time-dependent variation with multiple Davydov states, J. Chem. Phys. {\bf 146}, 124127 (2017).




\bibitem{Davydov1}A. S. Davydov, The Theory of Contraction of Proteins under their Excitation, J. Theor. Biol {\bf 38}, pp.559-569 (1973)


\bibitem{Davydov2}A. S. Davydov, and N. I. Kislukha, Solitary Excitons in One-Dimensional Molecular Chains, Phys. Status Solidi B {\bf 59}, pp.465-470 (1973)


\bibitem{Davydov4}A. Scott, Davydov's soliton, Phys. Rep. {\bf 217}, pp.1-67 (1992).


\bibitem{shor73} H.B.~Shore and L.M.~Sander, Ground State of the Exciton-Phonon System, {Phys. Rev. B}\textbf{7}, 4537 (1973).

\bibitem{jcppersp} Y.~Zhao, The hierarchy of Davydov's Ansaetze: from guesswork to numerically ``exact" many-body wave functions,  J. Chem. Phys. {\bf 158}, 080901 (2023).


\bibitem{Davydov5}N. Zhou, L. Chen, D. Mozyrsky, V. Chernyak, Y. Zhao,  Ground-state properties of sub-Ohmic spin-boson model with simultaneous diagonal and off-diagonal coupling, Phys. Rev. B Condens. Matter {\bf 90}, 155135 (2014)


\bibitem{Davydov6}N. Zhou, Z. Huang, J. Zhu, V. Chernyak, Y. Zhao, Polaron dynamics with a multitude of Davydov D2 trial states, J. Chem. Phys. {\bf 143}, 014113 (2015)


\bibitem{Davydov7}Y. Zhao, B. Luo, Y. Zhang, J. Ye, Dynamics of a Holstein polaron with off-diagonal coupling, J. Chem. Phys. {\bf 137}, 084113 (2012)


\bibitem{Davydov8}Y.Zhao, D. W. Brown, and K. Lindenberg, Variational energy band theory for polarons: Mapping polaron structure with the Toyozawa method, J. Chem. Phys. {\bf 107}, 3159 (1997)


\bibitem{QED_Huang} Z.~Huang, F.~Zheng, Y.~Zhang, Y.~Wei, and Y.~Zhao, Dissipative dynamics in a tunable Rabi dimer with periodic harmonic driving, \textit{J. Chem. Phys.} \textbf{150}, 184116 (2019).


\bibitem{QED_Zheng}  F.~Zheng, Y.~Zhang, L.~Wang, Y.~Wei, and Y.~Zhao, Engineering Photon Delocalization in a Rabi Dimer with a Dissipative Bath, \textit{Ann. Phys.} \textbf{530}, 1800351 (2018).

\bibitem{TMD} K.~Sun, K.~Shen, M.F.~Gelin, and Y.~Zhao, Exciton dynamics and time-resolved fluorescence in nanocavity-integrated monolayers of transition metal dichalcogenides, \textit{J. Phys. Chem. Lett.} {\bf 14}, 221-229 (2023).

\bibitem{Perspetive} Y. Zhao. The hierarchy of Davydov’s Ans\"atze: From
guesswork to numerically “exact” manybody wave functions. J. Chem. Phys. {\bf 158}, 080901 (2023)


\bibitem{D2-dw1}K. Sun, M. Gelin, Y. Zhao, Engineering Cavity Singlet Fission in Rubrene, J. Phys, Chem. Lett. {\bf 13}, 4090-4097 (2022)


\bibitem{D2-dw2}Z. Huang, M. Hoshina, H. Ishihara, Y. Zhao, Transient dynamics of super Bloch oscillations of a one dimensional Holstein polaron under the influence of an external AC electric field, Ann. Phys. {\bf 531}, 1800303 (2019)




\bibitem{D2-dw4}K. Sun, Q. Xu, L. Chen, M. Gelin, Y. Zhao, Temperature effects on singlet fission dynamics mediated by a conical intersection, J. Chem. Phys. {\bf 153}, 194106 (2020)


\bibitem{D2-dw5}Y. Fujihashi, L. Chen, A. Ishizaki, J. Wang, Y. Zhao, Effect of high-frequency modes on singlet fission dynamics, J. Chem. Phys. {\bf 146}, 044101 (2017)


\bibitem{D2-norm}Y. Yan, L. Chen, J. Luo, Y. Zhao, Variational approach to time-dependent fluorescence of a driven qubit, Phys. Rev. A {\bf 102}, 023714 (2020)


\bibitem{Core-Statis}A.D. Kammogne, M.B. Kenmoe and L.C. Fai, Statistics of interferograms in three-level systems, Phys. Lett. A {\bf 425}, 127872 (2022)


\bibitem{three1}G. Wang, D. Ye, L. Fu, X. Chen, J. Liu, Landau-Zener tunneling in a nonlinear three-level system, Phys. Rev. A {\bf 74}, 033414 (2006)



\bibitem{three2}M. B. Kenmoe, H. Phien, M. Kiselev, L. Fai, Effects of colored noise on Landau-Zener transitions: Two and Three-level systems, Phys. Rev. B {\bf 87}, 224301 (2013)


\bibitem{three3}M. B. Kenmoe, A. B. Tchapda, L. Fai,
SU(3) Landau-Zener interferometry with a transverse periodic drive, Phys. Rev. B {\bf 96}, 125126 (2017)

\bibitem{Qin16} X.-K. Qin.
How to control the coherent oscillations in Landau–Zener–Stueckelberg dynamics of three-level system,
Mod. Phys. Lett. {\bf 30}, 1650149 (2016)


\bibitem{CDT}F. Grossmann, Coherent Destruction of Tunneling, Phys. Rev. Lett. {\bf 67}, 516 (1991)

\bibitem{Malla18}R. K. Malla and M. E. Raikh.
Landau-Zener transition in a two-level system coupled to a single highly excited oscillator,
Phys. Rev.  B {\bf 97}, 035428 (2018)


\bibitem{Core-contour_plot}A. A. Boris and V. M. Krasnov, Quantization of the superconducting energy gap in an intense microwave field, Phys. Rev. B {\bf 92}, 174506 (2015)


\bibitem{CDT1}J. Gong, L. Morales-Molina, and P. H\"anggi, Many-Body Coherent Destruction of Tunneling, Phys. Rev. B Condens. Matter, {\bf 103}, 133002 (2009)



\bibitem{CDT2}G. Della Valle, M. Ornigotti, E. Cianci et al, Visualization of coherent destruction of tunneling in an optical double well system, Phys. Rev. Lett. {\bf 98}, 263601 (2007)


\bibitem{CDT3}S. Mukherjee, M. Valiente, N. Goldman et al, Observation of pair tunneling and coherent destruction of tunneling in arrays of optical waveguides, Phys. Rev. A {\bf 94}, 053853 (2016)

\bibitem{Gelin09} M. F. Gelin, D. Egorova, W. Domcke. Manipulating electronic couplings and nonadiabatic nuclear dynamics
with strong laser pulses. J. Chem. Phys. {\bf 131}, 124505 (2009)












\end{thebibliography}

\end{document}